\documentclass[usenatbib]{mnras}

\usepackage{newtxtext,newtxmath}

\usepackage[T1]{fontenc}

\DeclareRobustCommand{\VAN}[3]{#2}
\let\VANthebibliography\thebibliography
\def\thebibliography{\DeclareRobustCommand{\VAN}[3]{##3}\VANthebibliography}

\usepackage{graphicx}	
\usepackage{amsmath}	
\usepackage{float}
\usepackage[nolist]{acronym}
\usepackage{cuted}

\hypersetup{urlcolor=violet}

\usepackage{stfloats}
\usepackage{cuted}
\newcommand{\wideeq}[2]{
    \begin{figure}[b]
        \noindent\rule{\textwidth}{0.5pt} 
        \begin{equation}\label{#1}
        \begin{split}
            #2
        \end{split}
        \end{equation}
        \noindent\rule{\textwidth}{0.5pt} 
    \end{figure*}
}
\usepackage{stfloats}

\newcommand{\ECOSMOG}{{\textsc{ecosmog}}}

\newcommand{\EFTCAMB}{{\textsc{eft-camb}}}
\newcommand{\EFTRAMSES}{{\textsc{eft-ramses}}}
\newcommand{\HICLASS}{{\textsc{hi-class}}}

\newcommand{\RAMSES}{{\textsc{ramses}}}
\newcommand{\LCDM}{{$\Lambda$CDM}}
\newcommand{\HICOLA}{{\textsc{hicola}}}
\newcommand{\AREPO}{{\textsc{mg-arepo}}}
\newcommand{\MGGLAM}{{\textsc{mg-glam}}}

\newacro{AMR}{adaptive mesh refinement}
\newacro{BAO}{baryonic acoustic oscillation}
\newacro{CDM}{cold dark matter}
\newacro{CMB}{cosmic microwave background}
\newacro{CPL}{Chevallier-Polarski-Linder}
\newacro{csG}{cubic scalar Galileon}
\newacro{cvG}{cubic vector Galileon}
\newacro{DE}{dark energy}
\newacro{DESI}{Dark Energy Spectroscopic Instrument}
\newacro{DGP}{Dvali-Gabadadze-Porrati}
\newacro{DoF}{degree of freedom}
\newacro{EFT}{effective field theory}
\newacro{EFTN}{EFT Negative}
\newacro{EFTofDE}{effective field theory of dark energy}
\newacro{EFTP}{EFT Positive}
\newacro{EoM}{equation of motion}
\newacro{EoS}{equation of state}
\newacro{FLRW}{Friedmann–Lemaître–Robertson–Walker}
\newacro{GCCG}{generalised cubic covariant Galileon}
\newacro{GCG}{generalised cubic Galileon}
\newacro{GR}{general relativity}
\newacro{GW}{gravitational waves}
\newacro{IC}{initial conditions}
\newacro{ISW}{integrated Sachs-Wolfe}
\newacro{LSS}{large-scale structure}
\newacro{MG}{modified gravity}
\newacro{nDGP}{normal-branch DGP}
\newacro{QSA}{quasi-static approximation}
\newacro{sDGP}{self-acceleration branch DGP}
\newacro{WFA}{weak-field approximation}
\newacro{QCDM}{quintessence cold dark matter}
\newacro{cG}{cubic Galileon}

\title[EFT-RAMSES]{\textsc{eft-ramses}: a code to simulate the effective field theory of dark energy}

\author[N. O. Woodcock et al.]{
Nathaniel Ota Woodcock$^{1}$\thanks{E-mail: nathaniel.k.woodcock@durham.ac.uk (NOW)},
Sownak Bose$^{1}$,
Yunhao Gao $^{2,1,3}$,
Baojiu Li$^{1}$
\\\\
$^{1}$Institute for Computational Cosmology, Ogden Centre West, Department of Physics, Durham University, South Road, Durham DH1 3LE, UK\\
$^{2}$National Astronomical Observatories, Chinese Academy of Sciences, Beijing, 100101, China\\
$^{3}$School of Astronomy and Space Sciences, University of Chinese Academy of Sciences, Beijing, 100049, China}

\date{Accepted XXX. Received YYY; in original form ZZZ}

\pubyear{\the\year{}}

\begin{document}
\label{firstpage}
\pagerange{\pageref{firstpage}--\pageref{lastpage}}
\maketitle

\begin{abstract}
While the standard $\Lambda$CDM paradigm is in excellent agreement with most current cosmological observations, theoretical challenges surrounding the cosmological constant ($\Lambda$) have strongly motivated the exploration of
dynamical {\ac{DE}} and {\ac{MG}} models. Investigating the physical nature of the cosmic acceleration requires N-body simulations to probe the non-linear growth of cosmic structure {and prepare for the high-precision data from 
Stage-IV surveys}. In this paper, we present \EFTRAMSES, a comprehensive extension of the \ECOSMOG{} cosmological simulation code designed to explore non-linear structure formation in {\ac{DE} and} \ac{MG} scenarios. We embed the effective field theory of dark energy (EFTofDE) framework into this new numerical pipeline, utilising the $\alpha$-basis parameterisation to provide a versatile, model-agnostic, computational engine. By consolidating diverse scalar and vector-tensor theories—including the normal and self-accelerating \ac{DGP} models, cubic Galileons (\ac{csG}, \ac{cvG} and \ac{GCCG}), and generic \ac{EFT} parameterisation—into a single ``master'' Vainshtein equation, this pipeline bypasses the need for model-specific solvers {and easily specialises to any particular model}. As validations, we perform high-resolution N-body simulations for the \ac{nDGP}, \ac{csG}, \ac{GCCG}, and \ac{EFT} models, comparing the resulting matter power spectra against dependent and independent codes such as legacy \ECOSMOG{}, {\AREPO} and \HICOLA, as well as linear theory, and find excellent
agreement. \EFTRAMSES{} provides a robust and versatile computational tool for precise cosmological tests of \ac{DE} and \ac{MG} using upcoming cosmological surveys. The code is available for download from \href{https://github.com/nat-woodcock/EFT-Ramses}{the GitHub \EFTRAMSES{} repository}.

\end{abstract}




\section{Introduction}
The discovery of late-time cosmic acceleration \citep{Riess:1998cb,SupernovaCosmologyProject1999} fundamentally altered our understanding of the Universe, necessitating the introduction of a new energy component broadly termed `dark energy' (DE) to drive this accelerated expansion. Within the standard cosmological paradigm, this dark energy is represented by the cosmological constant, $\Lambda$, alongside {\ac{CDM}} in the \LCDM{} model. This model has proven to be an excellent empirical fit to a wide array of cosmological observations, including the {\ac{CMB}}, {\ac{BAO}}, and the {\ac{LSS}}, consistently supporting an equation of state parameter close to $w = -1$. Despite its immense observational success, the cosmological constant remains theoretically problematic; the vast discrepancy between its observed value and the theoretical prediction from quantum field theory lacks a compelling fundamental explanation. Furthermore, emerging observational results are beginning to challenge the static nature of $\Lambda$. Recent analyses from the {\ac{DESI}} \citep{DESI2025} hint at a mild preference for dynamical dark energy, showing deviations from a constant $\Lambda$ by $2.5\sigma - 3.9\sigma$ \citep{DESI:2024BAO}. These theoretical tensions and new observational hints have strongly motivated the field to explore alternative theoretical frameworks for cosmic acceleration, including models of dynamical {\ac{DE}} and theories of modified gravity ({\ac{MG}}).

Cosmology has entered a data-driven era, with Stage-IV surveys such as \ac{DESI}  \citep{DESICollaboration2023}, Euclid \citep{EuclidCollaboration2025}, and 
and the Vera C. Rubin Observatory's Legacy Survey of Space and Time (LLST) \citep{LSSTDESC2012} poised to map the \ac{LSS} of the Universe with unprecedented precision to investigate the physical nature of cosmic acceleration. Historically, standard \ac{DE} and \ac{MG} models have been distinguished by their distinct effects on the Universe. While \ac{DE} often affects \ac{LSS} solely by altering the background expansion rate, \ac{MG} models also directly influence matter clustering by modifying the underlying gravitational force. Consequently a degeneracy arises, since both models can produce identical background expansion histories. Therefore, it has been difficult to break the degeneracy between \ac{MG} and \ac{DE} using background expansion alone, as doing so strictly requires probing the non-linear growth of cosmic structure. 

Furthermore, previous studies have relied on phenomenological approaches \citep{Caldwelletal2007,Amendola2008} to quantify deviations from {\ac{GR}}, such as modifying the Poisson equation without deriving the modifications from a fundamental Lagrangian. In terms of the expansion history, another parameterisation is the {\ac{CPL}} parameterisation \citep[$w_0-w_a$,][]{ChevallierPolarski2001}, which assumes that the dynamical equation of state of dark energy $w_{\textrm{DE}}$ is a linear function of the scale factor $a$. Although these approaches have been popular, these parameterisations can introduce theoretical inconsistencies-such as ghost or gradient instabilities when predicting cosmological observables. To overcome this, the field requires a robust, unifying mathematical framework. The {\ac{EFTofDE}} \citep[][]{Gubitosi:2013,Bellini2014,Frusciante:2019xia} provides this consistency by systemically encompassing all single-scalar tensor theories, including the extensive Horndeski class \citep{Horndeski:1974wa}, into a single, model-agnostic action. Within this {\ac{EFT}} framework, the {generalised cubic Galileon} emerges as a highly compelling specific physical model to test \citep[][]{Deffayetetal2009}. The generalised cubic Galileon features kinetic braiding and a Vainshtein screening mechanism \citep{Will2014} that dynamically suppresses scalar fifth-force effects in high-density environments. This screening is essential, as it allows the model to satisfy stringent local gravity constraints while still actively enhancing structure formation on cosmic scales.


The \ac{EFTofDE} has emerged as a highly successful, unifying framework for testing departures from \ac{GR} on cosmological scales \citep{Gubitosi:2013,Bellini2014,Frusciante:2019xia}. This framework encapsulates all classes of single scalar-field theories, including the extensive Horndeski theory of gravity \citep{Horndeski:1974wa} into a single, model-independent gravitational action by parameterising the dynamics of \ac{DE} and \ac{MG} through expansion of operators that respect unbroken spatial diffeomorphisms. This theoretical unity eliminates the traditional need to derive and constrain the phenomenological equations for every individual theory of gravity. The rich phenomenological features of these models have been studied in the linear regime by using standard linear Einstein-Boltzmann solvers, primarily through public codes such as {\EFTCAMB} and {\HICLASS} \citep{Huetal2014,Zumalacarregui2017}. These codes have played a crucial role and have become the standard tools for constraining \ac{MG} parameters using \ac{CMB}  and \ac{LSS}  data \citep{Bellinietal2018}. These solvers have proven to be highly effective in the linear regime. However, fully exploiting the high-precision, non-linear information from upcoming stage-IV surveys requires extending the \ac{EFTofDE} framework into the non-linear regime and embedding it into high-resolution N-body simulations.

This work {involves} an extension to an  N-body simulation code developed to study cosmic structure formation under \ac{MG} scenarios within the \ac{EFTofDE} framework. We specifically focus on a subset of the Horndeski class of models, where an additional scalar {\ac{DoF}} couples with the spacetime metric.
To codify the \ac{EFTofDE}, our framework implements two parallel approaches for parameterising the system's Lagrangian: the phenomenological $\alpha$-basis, and the original \ac{EFT} parameterisation. By offering both bases simultaneously, our code provides enhanced versatility over existing implementations-which typically rely solely on the $\alpha$-basis \citep{Lietal2012}, allowing for a more direct and flexible mapping of specific theoretical models.

Implementing the Vainshtein screening mechanism within this framework presents a significant computational challenge. The {\ac{EoM}} for the additional scalar field is highly non-linear and involves products of spatial scalar field derivatives, rendering standard Fourier-based solvers and tree methods ineffective at small, highly clustered scales. To overcome this, we utilise a grid-based architecture equipped with robust non-linear multigrid solvers. Specifically, we introduce \EFTRAMSES, developed from the \RAMSES-based \ECOSMOG{} code \citep{Teyssier2002,Lietal2012}, which employs {\ac{AMR}} to dynamically achieve the high spatial resolutions required to accurately resolve these non-linear screening dynamics.

Ultimately, this new implementation embeds the \ac{EFTofDE} framework into our N-body pipeline, providing a powerful, generic engine for simulating a broad parameterisation of \ac{DE} and \ac{MG} models.  It is important to note that our current implementation explicitly operates under the quasi-static limit \citep{Zumalacarregui2017,Copeland2019,Bellini2016,Alonso2017,Gleyzes2016,Chudaykin2024}; consequently, operators that govern the clustering of dark energy—specifically $M_2^4$ (which dictates the dark energy sound speed) does not appear in the reduced field equations and are left for future. Despite this specific restriction, the framework remains exceptionally versatile, capable of capturing diverse non-linear dynamics and accommodating both Vainshtein and chameleon-type screening mechanisms. The primary advantage of this \ac{EFT}-based approach is its theoretical and computational generality. Rather than treating different gravity models as having structurally distinct field equations, this framework demonstrates that their non-linear screening dynamics are structurally similar, differing primarily by their time-dependent background coefficients. By mapping these theories into a single, unified ``master'' Vainshtein equation, our code naturally encompasses a wide array of popular theories - including the normal and self-accelerating branches of the {\ac{DGP}} model \citep{Dvalietal2000}, {\ac{csG}} and {\ac{cvG}} \citep{Barreira:2013,Barreiraetal2014}, and the {\ac{GCCG}} \citep{Ataydeetal2024,Fruscianteetal2020}. Because this master equation encapsulates the non-linear screening dynamics of multiple theories into one generalised format, we can restrict our \ac{EFT} parameters to perfectly mimic a wide array of \ac{MG} models such as the \ac{nDGP}, \ac{csG} and the \ac{GCCG} model, featured in this work, allowing us to directly cross-validate our new computational pipeline against existing, model-specific simulations to strictly verify its accuracy.

The remainder of this paper is organised as follows. In Section \ref{sect:2}, we introduce the theoretical framework of the \ac{EFTofDE} and explicitly derive the mapping of the generalised cubic Galileon model into this formalism, yielding both the background and perturbation equations. Section \ref{sect:individual_models} details the discrete numerical implementation within \EFTRAMSES{}, covering code units and showing how the models we tested can be unified into a single `master' Vainshtein equation. Section \ref{sect:eft-ramses} details our computational methodology, outlining the architecture of the \EFTRAMSES{} N-body code, explaining how its multigrid solver handles the highly non-linear scalar field dynamics, and presents the numerical validation of our pipeline against established baseline simulations. Finally, Section \ref{sect:discussion} provides a summary of our findings and conclusions.

Throughout this work, we adopt the 
metric signature $(-, +, +, +)$ for the spacetime metric $g_{\mu\nu}$ and operate in natural units where the speed of light is set to unity ($c = 1$). Greek indices ($\mu, \nu, \dots$) run over the spacetime coordinates $(0, 1, 2, 3)$, {while Roman indices ($i,j,\cdots$) run over the space coordinates $(1,2,3)$}. The canonical kinetic term of {a} scalar field {$\phi$} is defined using the metric as $X = g^{\mu\nu}\nabla_\mu\phi\nabla_\nu\phi$. {On the cosmological background we have $X=-\dot{\bar{\phi}}^2$ where an overdot denotes the time derivative of a quantity, unless otherwise stated. However, in what follows we will frequently ignore the overbar when there is no confusion, to lighten our notation.} Furthermore, we interchangeably use the gravitational couplings $8\pi G = \kappa = M_{\text{Pl}}^{-2}$, where $G$ is Newton's constant and $M_{\text{Pl}}$ the reduced Planck mass. We normally use a subscript $_0$ to denote the present-day value of a quantity. However, for the density parameter, such as $\Omega_{\text{DE}}$, we ignore this subscript and instead write the time dependence explicitly, e.g., $\Omega_{\text{DE}}(a)$ for their values at times other than today. The same rule applies to the dark energy \ac{EoS} parameter, with $w_{\text{DE}}(a)$ denoting the \ac{EoS} at arbitrary time while $w_{\text{DE}}$ its value today.

\section{Theory}\label{sect:2}

This section provides a 
{brief} overview of the {\ac{EFTofDE}}. We begin by introducing the generalised cubic Galileon model, followed by a derivation of the modified Poisson equation and the scalar field \ac{EoM} under the {\ac{QSA}}. These results are subsequently mapped into the \ac{EFTofDE} framework. 
This mapping translates the specific physical dynamics of the generalised cubic Galileon into the generalised language of \ac{EFTofDE}, establishing the formal mathematical structure required for numerical implementation in our N-body code.

\subsection{Horndeski Theory of Gravity}
\label{subsect:horndeski}

We start with the action of the Horndeski theory of gravity \citep{Horndeski:1974wa}, {which is an umbrella for single scalar field models with second-order EoM:}
\begin{equation}\label{eq:horndeski_action}
S_{\textrm{H}} = \int d^4x \sqrt{-g} \left[ L_2 + L_3 + L_4 + L_5 + L_m\right],
\end{equation}
where $g$ is the determinant of the metric tensor $g_{\mu\nu}$ and $L_m$ is the standard matter Lagrangian. The definitions of {the scalar field Lagrangian densities} $L_2$--$L_5$ are:
\begin{align}
L_2 &= K(\phi, X), \\
L_3 &= G_3(\phi, X)\Box\phi, \\
L_4 &= \frac{M_{\text{Pl}}^2}{2} G_4(\phi, X) R - M_{\text{Pl}}^2 G_{4X}(\phi, X)[(\Box\phi)^2 - \phi_{;\mu\nu}\phi^{;\mu\nu}] \,, \\
L_5 &= G_5(\phi, X)G_{\mu\nu}\phi^{;\mu\nu} \nonumber\\&\quad +\frac{1}{3}G_{5X}(\phi, X) \left[ (\Box\phi)^3 - 3\Box\phi\phi_{;\mu\nu}\phi^{;\mu\nu} + 2\phi_{;\mu\nu}\phi^{;\mu\sigma}\phi^{;\nu}_{\phantom{;\nu};\sigma} \right],
\end{align}
where $R$ is the Ricci scalar, $\phi$ is the scalar field representing the additional dynamical \ac{DoF}, $X\equiv g^{\mu\nu}\nabla_\mu\phi\nabla_\nu\phi$, $K(\phi,X)$ is its generalised kinetic term, $G_{3-5}(\phi,X)$ are {arbitrary functions of $\phi$ and $X$ that define the cubic, quartic and quintic-order scalar field self-interactions, respectively,} and $\Box\equiv\nabla^\mu\nabla_\mu$ represents the d'Alembert operator. We use a subscript $\phi$ ($X$) to represent the partial derivative with respect to $\phi$ ($X$), e.g., $G_{4\phi}\equiv\partial G_4/\partial\phi$, $G_{5X}\equiv\partial G_5/\partial X$.

We set $G_5=0$ and restrict $G_4$ dependence purely to the scalar field $\phi$ (i.e., $G_4=G_4(\phi)$ with $G_{4X} = 0$), to strictly enforce that the tensor propagation speed remains identical to the speed of light, $c$, consistent with the stringent observational bounds from GW170817 \citep{Abbott2017GW170817}. While the non-minimal coupling to gravity is explicitly preserved through this $\phi$-dependence, omitting the kinetic $X$-dependence is an observational necessity rather than a theoretical prerequisite for such coupling. Ultimately, this formulation encompasses all cosmologically viable theories allowed up to cubic order (generalised cubic Galileon), guaranteeing that the resulting \ac{EoM} remains strictly second-order while seamlessly accommodating 
{various} possible background expansion histories and non-linear screening mechanisms within this class of gravity models.

Varying the above action with respect to the metric and the scalar fields gives, respectively, the Einstein and scalar field equations:
\begin{equation}\label{eq:gcg_einstein}
    G_{\mu\nu} = \kappa \left[ T_{\mu\nu}^m + T_{\mu\nu}^{2} + T_{\mu\nu}^{3} +T_{\mu\nu}^{4} \right],
\end{equation}
\begin{align}\label{eq:gcg_scalar_eom}
0 =& \; K_{\phi} - 2K_{X}\Box\phi - 2K_{\phi X}X - 2K_{XX}\nabla_{\mu}X\nabla^{\mu}\phi \nonumber \\
& + 2G_{3\phi}\Box\phi + G_{3\phi\phi}X + 2G_{3\phi X}\nabla_{\mu}X\nabla^{\mu}\phi - 2G_{3\phi X}X\Box\phi \nonumber \\
& - 2G_{3X} \left[ (\Box\phi)^{2} - \nabla_{\mu}\nabla_{\nu}\phi\nabla^{\mu}\nabla^{\nu}\phi - R_{\mu\nu}\nabla^{\mu}\phi\nabla^{\nu}\phi \right] \nonumber \\
& - 2G_{3XX}\nabla_{\mu}X\nabla^{\mu}\phi\Box\phi + G_{3XX}\nabla_{\mu}X\nabla^{\mu}X+ \frac{M_{\text{Pl}}^2}{2} R \, G_{4\phi}.
\end{align}

{The scalar field's contribution to the total stress energy tensor is given by}:
\begin{equation}\label{eq:sf_stress_energy_tensor}
\begin{split}
& T_{\mu\nu}^{2} + T_{\mu\nu}^{3}+T_{\mu\nu}^{4} \\ 
=\,\, & g_{\mu\nu} K - 2 K_{X} \nabla_\mu \phi \nabla_\nu \phi -2 G_{3X} \Box \phi \nabla_\mu \phi \nabla_\nu \phi \\
& + 4 G_{3X} \nabla_{(\mu} \phi \nabla_{\nu)} \nabla_\alpha \phi \nabla^\alpha \phi - 2 g_{\mu\nu} G_{3X} \nabla^\alpha \phi \nabla^\beta \phi \nabla_\alpha \nabla_\beta \phi \\
& + 2 G_{3\phi} \nabla_\mu \phi \nabla_\nu \phi - g_{\mu\nu} G_{3\phi} (\nabla_\alpha \phi \nabla^\alpha \phi) + \frac{M_{\text{Pl}}^2}{2}G_4 G_{\mu\nu},
\end{split}
\end{equation}
{where} $G_{\mu\nu}$ is the Einstein tensor, and $T_{\mu\nu}^{m}$ represents the stress-energy tensor of all the other energy species in the universe. 

These equations represent the most general mathematical formulation of the generalised cubic Galileon model \citep{Deffayetetal2009}. Due to the fact that the Lagrangian components $K(\phi,X)$ and $G_3(\phi,X)$ are left as arbitrary functions of both the scalar field ($\phi$) and its kinetic term ($X$), this formalism avoids committing to a 
{specific} theory with a phenomenological potential or 
{particular} coupling strength. While this approach successfully maintains theoretical generality, systematically testing this vast array of arbitrary functions against observational data {poses a serious challenge, and} requires a more unified mathematical bridge. 

To this end, the field 
{has introduced} the \ac{EFTofDE}, which translates these complex scalar-tensor theories into a 
{unified}, model-agnostic language to 
{describe perturbation evolution.}

\subsection{The Effective Field Theory of Dark Energy}
\label{subsect:eftofde}
The effective field theory of dark energy (\ac{EFTofDE}) \citep{Gubitosi:2013,Bellini2014,Frusciante:2019xia} breaks the degeneracy between \ac{MG} and \ac{DE} to unify 
{the various} complex extensions {to \LCDM}. This model-independent framework encompasses most single-field extensions of \ac{GR}, which ensures theoretical consistency and characterises deviations from \ac{GR} at the level of cosmological perturbations.

The \ac{EFTofDE} 
constructs the most general action for linear perturbations around a {\ac{FLRW}} background, accommodating a single additional scalar \ac{DoF}. By formulating the theory in the unitary gauge \citep{Kodama:1984}, the \ac{EFTofDE} action is constructed using operators that preserve the residual symmetry of time-dependent spatial diffeomorphism. These operators are then systematically organised by their order in perturbations and spatial derivatives. Crucially, these residual symmetries allow each operator to be multiplied by an arbitrary time-dependent function, directly encoding any deviations from the standard {\LCDM} cosmology at the level of the action. Because of this broad parameterisation, the framework naturally permits the wide array of Horndeski theories, hence we included the most flexible, the generalised cubic Galileon models investigated herein. One can translate the model Eq.~\eqref{eq:horndeski_action} to the \ac{EFT} language, following the \cite{Gleyzes:2013ooa,Gleyzes2015Unifying} notation. 
In the unitary gauge this is written as:
\begin{equation}\label{eq:eft_parameterisation}
\begin{aligned}
    S_{EFT}=&\int d^4x\sqrt{-g}\bigg[\frac{M^2_*}{2}f(t)R-\Lambda(t)-c(t)g^{00}+\\&\frac{M^4_2(t)}{2}(\delta g^{00})^2-\frac{\bar{M}^3_1(t)}{2}\delta g^{00}\delta K-\frac{\bar{M}^2_2(t)}{2}(\delta K)^2...\bigg],
\end{aligned}
\end{equation}
where 
$\delta g^{00}$ represents the perturbation of the 
time-time metric component and $\delta K$ is the trace of the perturbed extrinsic curvature $K_{\mu\nu}$. The ellipsis denotes higher-order terms affecting perturbations, while $\{f(t), \Lambda(t), c(t), M^4_2(t), \bar{M}^3_1(t), \bar{M}_2^2(t), \dots \}$ are the free \ac{EFT} functions. Note that the overbar in $\bar{M}_1(t)$ and $\bar{M}_2(t)$ widoes not denote the background average here, but rather used as a label to distinguish the operators that couple to the extrinsic curvature (such as $\delta K$ and its trace) from the unbarred operators (like $M_2^4(t)$) that couple to the metric perturbations (such as $\delta g^{00}$). From \cite{Gleyzes:2013ooa,Gleyzes2015Unifying}, the relations between the \ac{EFT} parameters and $K$, $G_3$ and $G_4$ in the $S_{\text{EFT}}$ action can be specified as:
\begin{equation}\label{eq:eft_horndeski_mapping}
\begin{aligned}
    f(t) &= G_4(\phi), \\
\Lambda(t) &= \dot{\phi}^2 G_{3X}(\phi, X)(3H\dot{\phi} + \ddot{\phi}) - K_X(\phi, X)\dot{\phi}^2 - K(\phi, X), \\
c(t) &= \dot{\phi}^2 G_{3X}(\phi, X)(3H\dot{\phi} - \ddot{\phi}) + \dot{\phi}^2 G_{3\phi}(\phi, X) - K_X(\phi, X)\dot{\phi}^2, \\
M_2^4(t) &= \frac{1}{2}\dot{\phi}^2 G_{3X}(\phi, X)(3H\dot{\phi} + \ddot{\phi}) - 3H\dot{\phi}^5 G_{3XX}(\phi, X)\\
&\quad - \frac{1}{2}\dot{\phi}^4 G_{3\phi X}(\phi, X)  + K_{XX}(\phi, X)\dot{\phi}^4, \\
\bar{M}_1^3(t) &= -2\dot{\phi}^3 G_{3X}(\phi, X),
\end{aligned}
\end{equation}
where $H$ is the Hubble parameter, and we have used the background values of $\phi$ and $X$ \footnote{\label{ft1}{We usually use an overbar for the background value of a quantity, but here neglect it to lighten our notation. The context should make it clear that these equations contain only the background values of $\phi$ and $X$.}}. In this paper, the overdot represents the derivative with respect to the cosmic time $t$.

A key strength of the \ac{EFTofDE} is its ability to decouple the background expansion history from the evolution of cosmic perturbations. The background evolution can be specified to reproduce \LCDM{} or arbitrary dynamical dark energy models, while departures from \ac{GR} are systematically parameterised by time-dependent functions governing the scalar and tensor modes. This clear separation provides a powerful tool for testing deviations from both \LCDM{} and \ac{GR} with cosmological observables.  On the background, the first three parameters in the action are related as:
\begin{equation}\label{eq:horndeski_action4}
\begin{aligned}
c &= \frac{1}{2} (\rho_{\text{DE}} + P_{\text{DE}}) + \frac{1}{2} \left( -\ddot{f} + H \dot{f} \right) M_{\text{Pl}}^2, \\
\Lambda &= \frac{1}{2} (\rho_{\text{DE}} - P_{\text{DE}}) + \frac{1}{2} \left( \ddot{f} + 5H \dot{f} \right) M_{\text{Pl}}^2,
\end{aligned}
\end{equation}
where $\rho_{\text{DE}}$ and $P_{\text{DE}}$ are the density and pressure of \ac{DE} respectively. 

To characterise linear perturbations, it is widely popular to introduce four time-dependent, phenomenological functions \citep{Bellini2014}, dubbed $\alpha$-basis. These dimensionless functions are denoted as $\alpha_{\textrm{M}}$, $\alpha_{\textrm{K}}$, $\alpha_{\textrm{B}}$ and $\alpha_{\textrm{T}}$ which represent specific physical properties of the Horndeski theory and efficiently encapsulate the effects of \ac{DE} and \ac{MG}. Here, $\alpha_{\textrm{M}}$ is the running Planck mass, which measures the time evolution of the effective Planck mass. $\alpha_{\textrm{B}}$ is the braiding term, which represents the mixing of the \ac{DE} field and the metric. $\alpha_{\textrm{K}}$ is known as the kinetic term, which affects the propagation speed of the \ac{DE} field, and $\alpha_{\textrm{T}}$ is known as the tensor speed excess and defines how {\ac{GW}} deviate in their propagation speed from the speed of light. Presently, this formulation has proven to be indispensable for Einstein-Boltzmann solvers and N-body simulations, successfully bridging theoretical models with cosmological observables. To characterise linear scalar perturbations within Horndeski-type theories \citep{Horndeski:1974wa}, this action is typically mapped onto four dimensionless $\alpha$-basis functions \citep{Bellini2014,Gleyzes2015Unifying,Gleyzes2015Horndeski,Frusciante2016,Amendola2018}:
\begin{equation}\label{eq:horndeski_action5}
\begin{aligned}
\alpha_{\text{K}} &\equiv \frac{2c + 4M_2^4}{M^2 H^2},\\ \alpha_{\text{B}} &\equiv \frac{M_{\text{Pl}}^2 \dot{f} - \bar{M}_1^3}{2M^2 H},\\
\alpha_{\text{M}} &\equiv \frac{M_{\text{Pl}}^2 \dot{f} + 2(\bar{M}^2_2)^{\cdot}}{M^2 H},\\ 
\alpha_{\text{T}} &\equiv -\frac{2\bar{M}^2_2}{M^2},
\end{aligned}
\end{equation}
where $M$ is the time-dependent effective Planck mass. While the full basis includes the kinetic term ($\alpha_{\textrm{K}}$) and tensor speed excess ($\alpha_{\textrm{T}}$), modern analyses primarily focus on the running Planck mass ($\alpha_{\textrm{M}}$) and the braiding term ($\alpha_{\textrm{B}}$). These parameters capture the time evolution of the effective gravitational coupling and the kinetic mixing between the \ac{MG} field and the metric, respectively. In our work, we chose $\alpha_{\textrm{T}}=0$ due to the detections of GW170817 and its electromagnetic counterpart arriving simultaneously \citep{Abbott2017GW170817}, which has set a strong constraint on this parameter $a_T<10^{-15}$ \citep{Baker2017,CreminelliVernizzi2017,EzquiagaZumalacarregui2017} showing that it is negligible. More interestingly, $\alpha_{\textrm{K}}$ is proportional to the parameter $M_2^4$ that governs the clustering of \ac{DE}, which does not appear in our equation of motion and cosmological data have been shown to be insensitive to this parameter in the quasi-static limit \citep{Zumalacarregui2017,Copeland2019,Bellini2016,Alonso2017,Gleyzes2016,Chudaykin2024}. Therefore, our code does not support models that involve the clustering of \ac{DE}.

In the non-linear regime, describing \ac{MG} faces a strict dichotomy, which involves either systematically adding higher-order terms yielding unmanageable complexity, or utilising the background and linear observations to reconstruct the functional form of \ac{MG} Lagrangians. An abundance of models has been studied in this manner \citep{Gao:2025}. However, reconstructions of this type have been limited to a specific \ac{MG} theory, consequently coming at the cost of giving up generality. When extending our framework to these non-linear scales, we must consider the specific screening mechanisms that suppress deviations from GR. In this work, we directly map the generalised cubic Galileon to the \ac{EFTofDE} to capture its intrinsic Vainshtein screening. This mechanism is dynamically driven by the non-linear self-interactions of the scalar field—specifically the quadratic spatial derivative terms $(\partial^2\delta\phi)^2 - (\partial_i\partial_j\delta\phi)^2$. Furthermore, we isolate a targeted, minimal subset of the \ac{EFTofDE} framework that retains the primary non-linear interactions necessary for accurate cosmological N-body simulations. Consequently, our simulations successfully recover \ac{GR} in high-density environments, ensuring strict compatibility with local gravity tests.

The primary conceptual advantage of \ac{EFTofDE} is its theoretical completeness. Because it originates from a single covariant action, dynamical constraints and conservation laws are inherently respected. Furthermore, it allows for the strict imposition of physical viability conditions, ensuring the explored parameter space is free from ghost and gradient instabilities \citep{Gleyzes:2013ooa,Lombriser:2019jhb}. This allows observational constraints to be mapped directly onto the underlying field dynamics of specific scalar-tensor theories, such as the cubic Galileon models \citep{Barreira:2013,Frusciante:2019xia,Gao:2025}.

\subsection{Equations that govern structure formation}

To obtain the 
perturbation equations from Eq.~\eqref{eq:gcg_einstein} and Eq.~\eqref{eq:gcg_scalar_eom}, we use the perturbed {\ac{FLRW}} metric in the Newtonian gauge:
\begin{equation}\label{eq6}
    \text{d}s^2 = -(1 + 2\Phi)\text{d}t^2 + a^2(t)(1 - 2\Psi)\delta_{ij}\text{d}x^i\text{d}x^j,
\end{equation}
where $a$ is the scale factor, {$\Phi,\Psi$ the Bardeen potentials} and $\delta_{ij}$ the Kronecker delta representing the spatial part of the metric. 
$\phi$, 
$\Psi$, $\Phi$ are functions of time and space, 
{and $\phi$ can be split} as $\phi(t, \vec{x}) = \bar{\phi}(t) + \delta\phi(t, \vec{x})$, where an overbar represents the background averaged value of a quantity and $\delta\phi$ is the field perturbation.


{In a cosmological background,} the Friedmann equation and scalar field \ac{EoM}, Eq.~\eqref{eq:gcg_scalar_eom}, are given as follows:
\begin{equation}\label{eq:friedmann_eqn}
       3\left(1- \frac{G_4}{2}\right)H^2  = \frac{1}{M_{\text{Pl}}^2}\rho_m - K - 2K_X \dot{\phi}^2 + 6H G_{3X} \dot{\phi}^3 + G_{3\phi}\dot{\phi}^2,
\end{equation}
and
\begin{equation}\label{eq:scalar_eom_background}
\begin{aligned}
    \left[ -2G_{3\phi} + 2K_X + (2G_{3\phi X} - 4K_{XX})\dot{\phi}^2 + 12H\dot{\phi}(\dot{\phi}^2G_{3XX}-G_{3X}) \right] \ddot{\phi}\\ - 6H\dot{\phi}(G_{3\phi} - K_X + G_{3\phi X}\dot{\phi}^2) + (2K_{\phi X} - G_{3\phi\phi})\dot{\phi}^2\\ - 6G_{3X}\dot{\phi}^2(\dot{H} - H^2) + K_\phi + 6 (\dot{H} + 2H^2) G_{4\phi}= 0.
\end{aligned}
\end{equation}
Note that we have again neglected the overbar for background quantities. Together, Eq.~\eqref{eq:friedmann_eqn} and Eq.~\eqref{eq:scalar_eom_background} establish the complete, coupled background dynamics governing the cosmic expansion. 

The Quasi-Static Approximation (QSA) serves as a robust approximation for the Galileon model deep within the sub-horizon regime \citep{Barreiraetal2012,DeFeliceTsujikawa2011}. The core premise of the QSA is that the spatial derivatives of any perturbed quantity, $\delta Q$, dominate over its respective time derivatives, but do not necessarily dominate over the time derivative of background quantities. Explicitly, we assume $|\partial_i\delta Q| \gg |\dot{\delta Q}| \sim |H \delta Q|$ {and $|\partial^2\delta Q|\gg H|\partial_i\delta Q|$}. As a result, in the field equations, we neglect all the time derivatives of the perturbed quantities (e.g., $\dot{\delta\phi}$, $\dot{\Psi}$) and preserve only the terms featuring the highest number of spatial gradients--specifically, linear terms like $\partial^2\delta\phi$ and $\partial^2\Psi$, alongside dominant non-linear terms such as $(\partial^2\delta\phi)^2$. 

While first spatial derivatives ($\partial_i\Phi$ and $\partial_i\Psi$) must be retained in equations governing fluid dynamics, such as the Euler equation, they are heavily suppressed relative to the second spatial derivatives within the \ac{MG} field equations. Because these quantities are negligible on the scales of interest, we can safely apply linearising simplifications, such as $(1 - 2\Phi)\partial^2\delta\phi \simeq \partial^2\delta\phi$ {and $(\partial_i\delta\varphi)^2\ll\partial^i\partial_i\delta\phi$}. {These are known as weak-field approximations (WFA).} It is important to note that because the QSA explicitly drops certain linear time-derivative terms, its outputs cannot be directly compared to standard linear perturbations on very large, super-horizon scales. Instead, the validity of employing the combined QSA and WFA for cosmic structure formation has been firmly established by previous counterpart works that successfully benchmarked these approximations against full, un-approximated dynamical simulations.

Under these approximations, the modified Poisson equation and the anisotropy equation obtained through the (0,0) and the traceless $(i, j)$ component of equation \eqref{eq:gcg_einstein}, respectively, and the Galileon scalar field equation of motion \eqref{eq:gcg_scalar_eom} are given below:
\begin{equation}\label{eq9}
    \nabla^2\Phi = \frac{a^2}{2 M_{\text{Pl}}^2 G_4}\delta\rho_m + \frac{1}{M_{\text{Pl}}^2 G_4}\left( -\frac{M_{\text{Pl}}^2}{2} G'_4 - \dot{\phi}^2 G_{3X} \right) \nabla^2\delta\phi \,,
\end{equation}

\begin{equation}\label{eq:horndeski_action01}
    \nabla^2\Psi = \frac{a^2}{2 M_{\text{Pl}}^2 G_4}\delta\rho_m + \frac{1}{M_{\text{Pl}}^2 G_4}\left( +\frac{M_{\text{Pl}}^2}{2} G'_4 - \dot{\phi}^2 G_{3X} \right) \nabla^2\delta\phi \,,
\end{equation}

\begin{equation}\label{eq:horndeski_action0}
\begin{split}
    &\left[ -K_X + G_{3\phi} + 2G_{3X}(\ddot{\phi} + 2H\dot{\phi}) + \dot{\phi}\dot{G}_{3X} \right] \nabla^2\delta\phi \\
    &\qquad - \frac{G_{3X}}{a^2} \left[ (\nabla^2\delta\phi)^2 - (\partial_i\partial_j\delta\phi)^2 \right] \\
    &\quad = -\frac{M_{\text{Pl}}^2}{2} G_{4\phi}\nabla^2(2\Psi - \Phi) - G_{3X}\dot{\phi}^2\nabla^2\Phi \,.
\end{split}
\end{equation}
where $\delta\rho_m$ is the matter density perturbation { and $\nabla^2 = \partial_i \partial^i$ is the spatial Laplacian operator} and $'$ represents the derivative with respect to $\phi$.

We can combine these equations to eliminate the potentials $\Phi$ and $\Psi$ to obtain the N-body equation\footnote{{For simplicity we refer to the equation that is implemented in the N-body simulation code as the N-body equation.}}: 

\begin{strip}
\begin{equation}\label{eq:nbody_eqn_horndeski}
    \left[ \frac{3(G'_4)^2 - 2G'_4G_{3X}\dot{\phi}^2 - G_{3X}^2\dot{\phi}^4}{2G_4} - K_X + G_{3\phi} + 2G_{3X}(\ddot{\phi} + 2H\dot{\phi}) + \dot{\phi}\dot{G}_{3X} \right] \nabla^2\delta\phi - \frac{G_{3X}}{a^2} \left[ (\nabla^2\delta\phi)^2 - (\partial_i\partial_j\delta\phi)^2 \right] = -\frac{a^2}{4G_4} \left( G'_4 + G_{3X}\dot{\phi}^2 \right) \delta\rho_m,
\end{equation}
where $(\partial_i \partial_j \delta\phi)^2 = (\partial_i \partial_j \delta\phi)(\partial^i \partial^j \delta\phi)$. This equation represents the most general N-body equation when $M^2_4$ is not needed. 
{For completeness and convenience,} we convert Eq.~(\ref{eq:nbody_eqn_horndeski}) with the mappings in Eq.~(\ref{eq:eft_horndeski_mapping}) to obtain the N-body equation in the \ac{EFT} form:
\begin{equation}\label{eq:nbody_eqn_eft}
    \left[ \frac{3(M_{\text{Pl}}^2 \dot{f})^2 - 2(M_{\text{Pl}}^2 \dot{f})\bar{M}_1^3 - (\bar{M}_1^3)^2}{4\dot{\phi}^2 M_{\text{Pl}}^2 f} + \frac{1}{\dot{\phi}^2} \left( c(t) + \frac{1}{2}\dot{\bar{M}}_1^3 + \frac{1}{2}H \bar{M}_1^3 \right) \right] \nabla^2\delta\phi - \frac{\bar{M}_1^3}{2a^2\dot{\phi}^3} \left[ (\nabla^2\delta\phi)^2 - (\partial_i\partial_j\delta\phi)^2 \right] = -\frac{a^2}{4M_{\text{Pl}}^2 f \dot{\phi}} \left( M_{\text{Pl}}^2 \dot{f} + \bar{M}_1^3 \right) \delta\rho_m,
\end{equation}
and in the $\alpha$-basis form:
\begin{equation}\label{eq:nbody_eqn_alpha}
    \frac{1}{\dot{\phi}^2} \left[ \frac{1}{2}(\rho_{\text{DE}}+ P_{\text{DE}}) + M^2 \left( H^2(1+\alpha_{\text{B}})(\alpha_{\text{M}} - \alpha_{\text{B}}) - \dot{H}\alpha_{\text{B}} - H\dot{\alpha}_{\text{B}} \right) \right] \nabla^2\delta\phi- \frac{M^2 H(\alpha_{\text{M}} - 2\alpha_{\text{B}})}{2a^2\dot{\phi}^3} \left[ (\nabla^2\delta\phi)^2 - (\partial_i\partial_j\delta\phi)^2 \right] = - \frac{a^2 H}{2\dot{\phi}} (\alpha_{\text{M}} - \alpha_{\text{B}}) \delta\rho_m
\end{equation}
\end{strip}
\noindent{where we have used} several useful relations. 
By setting the tensor speed excess to zero ($\alpha_{\textrm{T}} = 0$), we can combine the $\alpha$-basis parameters ($\alpha_{\textrm{M}}$ and $\alpha_{\textrm{B}}$) to transform the \ac{EFT} parameters ($\bar{M}_1^3$, $c$, and $M_{\text{Pl}}^2 \dot{f}$) present in the overarching N-body equation:
\begin{equation}\bar{M}_1^3 = M^2 H (\alpha_{\text{M}} - 2\alpha_{\text{B}}) ,
\end{equation}
\begin{equation} c = \frac{1}{2}(\rho_{\text{DE}} + P_{\text{DE}}) + \frac{1}{2} M^2 \left[ H^2 \alpha_{\text{M}} (1 - \alpha_{\text{M}}) - \dot{H} \alpha_{\text{M}} - H \dot{\alpha}_{\text{M}} \right] ,
\end{equation}
\text{and}
\begin{equation}
\frac{M_{\text{Pl}}^2 \dot{f}}{2}= \frac{1}{2} M^2H \alpha_{\text{M}}.
\end{equation}
These relations successfully map the \ac{EFT} parameters to the $\alpha$-basis definitions, which will be utilised in Section \ref{sect:individual_models} to formulate the final discrete variables compliant with the {\ECOSMOG} code conventions.

An important implicit assumption in deriving the above equations is that the functions $K$, $G_3$ and $G_4$ are not strongly nonlinear, so all the terms in the coefficients can be taken as their corresponding background values. In these types of simulations, the scalar field perturbation $\delta\phi$ is generally of the same order as the potential $\Phi$, and $|\delta\phi|\ll|\bar{\phi}|$, and to a good approximation we have, e.g., $G_{3\phi}(\phi)\approx G_{3\phi}(\bar{\phi})$. This approximation can break down for highly nonlinear forms of $G_3(\phi)$, e.g., $G_3(\phi)=\phi^{1/\epsilon}$ with $0<\epsilon\lesssim|\delta\phi/\bar{\phi}|$. This corresponds to clustering dark energy, in which the scalar field has significant clustering: $\delta\rho_{\text{DE}}/\delta_{\text{DE}}$ can become $\mathcal{O}(1)$ or larger. In these cases the sound speed of the scalar field $c_{\text{s}}$ is generally much smaller than the speed of light, $c$, and, contrary to Eqs.~(\ref{eq:nbody_eqn_eft}-\ref{eq:nbody_eqn_alpha}), the N-body equations can no longer be fully determined by the \ac{EFT} functions or by the $\alpha$-basis functions. In particular, $M_2^4$, which determines $c_{\text{s}}$, will no longer be negligible and this can lead to a large variety of clustering dark energy models which cannot be simulated using the code developed here. A detailed study of this new class of models is beyond the scope of this work.

\section{Vainshtein screening Models}
\label{sect:individual_models}

In this section, we demonstrate how 
a wide array of Horndeski-type theories studied in the literature can be unified into a single ``master" Vainshtein equation, which simplifies the numerical implementation of diverse \ac{MG} models.

\subsection{Code Units and the Master Equation}
\label{subsect:code_units}

To evaluate the field equations numerically, we transform all physical variables into a system of dimensionless code units. This conversion adapts the 
supercomoving coordinate framework utilised in {\RAMSES} \citep{Teyssier2002,Lietal2012}, 
by the following definitions of code units:
\begin{equation}
\begin{aligned}
\tilde{\boldsymbol{x}} &= \frac{\boldsymbol{x}}{B}, & \tilde{\rho} &= \frac{\rho a^3}{\rho_c \Omega_m}, & \tilde{\boldsymbol{v}} &= \frac{a\boldsymbol{v}}{BH_0}, \\
\tilde{\Phi} &= \frac{a^2 \Phi}{(BH_0)^2}, & \text{d}\tilde{t} &= H_0 \frac{\text{d}t}{a^2}, & \tilde{\varphi} &= \frac{c^2 a^2 \varphi}{(BH_0)^2},
\end{aligned}
\end{equation}
{where} $\boldsymbol{x}$ denotes the comoving coordinate, $\boldsymbol{v}$ the particle velocity, and $\Phi$ is the standard Newtonian potential. 
The background cosmology is described by the present-day critical density $\rho_c$ and fractional matter energy density $\Omega_m$. To scale the physical quantities into code units, we use $B$, the comoving size of the simulation box in units of $h^{-1}\text{Mpc}$, alongside $H_0$, the Hubble constant expressed in $100h\,\text{km s}^{-1}\text{Mpc}^{-1}$.

By adopting this specific normalisation, all resulting transformed variables are strictly dimensionless, and the 
average matter density is conveniently fixed to unity ($\bar{\tilde{\rho}} = 1$). For the \ac{EFTofDE}, 
we apply an initial field redefinition to render the scalar field perturbation dimensionless:
\begin{equation}
    \varphi \equiv \frac{\delta\phi}{M_{\text{Pl}}}.
\end{equation}

{Using the above definitions, the scalar field \ac{EoM} and the modified Poisson equation can be written as the following unified form:
\begin{equation}\label{eq:scalar_eom_code_units}
    \nabla^2\varphi + \frac{R_c^2}{3\beta a^4} \left[ (\nabla^2\varphi)^2 - \nabla^i\nabla^j\varphi\nabla_i\nabla_j\varphi \right] = \frac{1}{\beta}\Omega_m a\delta,
\end{equation}
\begin{equation}\label{eq:poisson_eqn_code_units}
    \nabla^2\Phi = \frac{3}{2}\Omega_m a\delta + \alpha \nabla^2\varphi.
\end{equation}
where all quantities are in code units, although we have omitted their tildes for brevity. $\alpha,\beta$ and $R_c$ are functions of the background scalar field and hence time, the exact forms of which depend on the model being considered. This expression for Eq.~\eqref{eq:scalar_eom_code_units} takes exactly the same form as the \ac{DGP} model, which is convenient as described below. We will refer to Eq.~\eqref{eq:scalar_eom_code_units} as the \textit{master Vainshtein equation}, or the \textit{master equation} for brevity, from here on.}

The three time-dependent dimensionless functions, $\alpha$, $\beta$ and $R_c^2$, entirely encapsulate the distinct physical characteristics and nonlinear dynamics of all the \ac{MG} models of interest. By isolating the model-specific physics into these background functions, our new code, \EFTRAMSES, which will be described in more details in the next section, only requires a single and robust numerical solver for the wide variety of \ac{MG} models featuring nonlinear derivative couplings. This makes it a highly convenient and versatile computational pipeline. 

{In the linear perturbation regime, structure formation is governed by an effective Newton constant $G_{\mathrm{eff}}$, given by:
\begin{equation}
    \frac{G_{\textrm{eff}}}{G} = 1+\frac{2\alpha}{3\beta}.
\end{equation}
$R_c$, which in Eq.~\eqref{eq:scalar_eom_code_units} only appears in front of the nonlinear terms, does not contribute at linear order.}


\subsection{Specific Vainshtein Screening Models}
\label{subsect:specific_models}

{Models with the Vainshtein screening mechanism \citep{Vainshtein} are extremely versatile, originating from both general considerations of scalar-tensor (Horndeski) theories and considerations of extra spatial dimensions. These theories feature the $G_3$ term in the Horndeski action, Eq.~\eqref{eq:horndeski_action}, or the $\bar{M}_1^3(t)$ term in the \ac{EFTofDE} parameterisation, Eq.~\eqref{eq:eft_parameterisation}, thereby occupying a substantial portion of the viable theory space. They also have rich phenomenology.}

\subsubsection{The \ac{sDGP} model}
\label{subsubsect:sDGP}

The \ac{DGP} model \citep{Dvalietal2000} is a braneworld theory which posits that fields in the standard model of particle physics are confined to a four-dimensional ``brane'' embedded within an infinite, five dimensional Minkowski bulk. The gravitational dynamics of the universe are governed by a competition between the 5D Einstein-Hilbert action of the bulk and a 4D ``induced'' gravity term on the brane:
\begin{equation}\label{eq:dgp_action}
S_\textrm{DGP} = \int{\mathrm{d}}^5x \sqrt{-g^{(5)}} \left[ \frac{M_5^3}{2} R^{(5)} \right]
+ \int{\mathrm{d}}^4x \sqrt{-g} \left[ \frac{M_{\mathrm{Pl}}^2}{2} R + \mathcal{L}_m \right],
\end{equation}
where {$M_5$ is the 5D reduced Planck mass, $g^{(5)}$ is the determinant of the metric describing the 5D bulk, $R^{(5)}$ the Ricci scalar of the bulk}, and $\mathcal{L}_m$ represents the standard matter Lagrangian {density}. 

In this framework, gravity behaves according to standard \ac{GR} on small scales but ``leaks" into the bulk at large cosmological distances, a transition 
characterised by the crossover scale $r_c \simeq M_{\mathrm{Pl}}^2 / 2M_5^3$.
In the decoupling limit, the bending of the brane into the extra dimension manifests mathematically as an additional scalar degree of freedom, $\phi$, governed by an effective 4D Galileon Lagrangian {density}:
\begin{equation}\label{eq:lagdgp}
\mathcal{L}_{\phi} = -\frac{1}{2} \nabla_\mu \phi \nabla^\mu \phi \pm \frac{1}{2\Lambda_s^3} (\nabla_\mu \phi \nabla^\mu \phi) \Box \phi,
\end{equation}
where $\Lambda_s$ is the strong coupling energy scale. This 
Lagrangian maps seamlessly into {the} generalised Horndeski framework, isolating the dynamics to a standard kinetic term $K(\phi,X) \propto X$ and cubic derivative self-interaction $G_3(\phi,X) \propto X \Box\phi$. Most importantly, it is this 
cubic term that corresponds to the Vainshtein screening mechanism within local, high-density environments such as our solar system.

The $\pm$ sign in the effective Lagrangian {density}, Eq.~\eqref{eq:lagdgp}, explicitly separates the theory into its two distinct cosmological branches: the normal branch ({nDGP}) and the self-accelerating branch (\ac{sDGP}). The self-accelerating branch corresponds to the negative sign; this branch spontaneously produces late-time cosmic acceleration without the need for a cosmological constant, albeit at the theoretical cost of introducing ghost instabilities. For the \ac{sDGP} model, the background parameters governing our master Vainshtein equation are given by:
\begin{align}
  \label{eq:sDGP_beta} \beta &= 1 - \frac{E}{\sqrt{\Omega_{\text{rc}}}} \left( 1 + \frac{E'}{3E} \right) = - \frac{\frac{1}{2}\Omega_m a^{-3} + \Omega_{\text{rc}}}{\sqrt{{\Omega_{\text{rc}}} (\Omega_m a^{-3} + \Omega_{\text{rc}})}},\\
  \label{eq:sDGP_alpha} \alpha &= \frac{1}{2},\\
  \label{eq:sDGP_Rc} R_c^2 &= \frac{1}{4\Omega_{\text{rc}}},
\end{align}
where $E\equiv H/H_0$, $'\equiv\textrm{d}/\textrm{d}\ln(a)$ and $\Omega_{\textrm{rc}}\equiv1/(4H_0^2r_c^2)$ is the model parameter with $r_c$ being the 
crossover scale {introduced above}. The background expansion history is given as:
\begin{equation}\label{eq:sDGP_background}
    E = \sqrt{\Omega_ma^{-3}+\Omega_{\textrm{rc}}} + \sqrt{\Omega_{\textrm{rc}}}.
\end{equation}
{Eqs.~(\ref{eq:sDGP_beta}-\ref{eq:sDGP_background}) are all what is needed to run an N-body simulation for the \ac{sDGP} model. Note that here and below all the density parameters are set to their values today.}

\subsubsection{The \ac{nDGP} model}
\label{subsubsect:nDGP}

The normal branch of the \ac{DGP} model originates from the 
{same} 5D braneworld action, {Eq.~\eqref{eq:dgp_action},} but corresponds to the positive sign in the effective 4D 
Lagrangian {density, Eq.~}\eqref{eq:lagdgp}. Although this branch does not self-accelerate---meaning an additional smooth \ac{DE} component or a 
cosmological constant $\Lambda$ must be introduced to drive the observed late-time {acceleration}---it is theoretically 
{more sound} as it 
avoids the ghost instabilities that affect the self-accelerating branch. 

As the scalar field interactions mathematically mirror those of \ac{sDGP}, they obey the same structural Vainshtein screening mechanics, differing only by the background coefficients. For the nDGP model, the 
{master equation} parameters are given as:
\begin{align}
  \label{eq:nDGP_beta} \beta &= 1+\frac{E}{\sqrt{\Omega_{\textrm{rc}}}}\left(1+\frac{E'}{3E}\right),\\
  \label{eq:nDGP_alpha} \alpha &= \frac{1}{2},\\
  \label{eq:nDGP_Rc} R_c^2 &= \frac{1}{4\Omega_{\textrm{rc}}}. 
\end{align}
As this model does not {have} self-acceleration, the accelerated expansion has to be driven by something else. Two possibilities are often considered: in \ac{nDGP} with \LCDM{} background, this is assumed to be a smooth (i.e., no spatial clustering) dark energy component, labelled as `Q', so that the background expansion history is identical to that of flat \LCDM{} with the same $\Omega_m$. In this case we have: 
\begin{equation}
    \beta = 1 + \frac{\frac{1}{2}\Omega_ma^{-3}+\Omega_\Lambda}{\sqrt{\Omega_{\textrm{rc}}\left(\Omega_ma^{-3}+\Omega_\Lambda\right)}},
\end{equation}
and the background expansion history is given as: 
\begin{equation}
    E = \sqrt{\Omega_ma^{-3}+\Omega_\Lambda},
\end{equation}
with $\Omega_\Lambda\equiv 1-\Omega_m$.

The second possibility is where the accelerated expansion is driven by a cosmological constant, $\Lambda$. In this case, $\Omega_\Lambda$ is a free parameter, {and \ac{DE} is contributed by both it and the scalar field:}
\begin{equation}
    \Omega_m + \Omega_{\textrm{DE}} = \Omega_m + \Omega_\Lambda + \Omega_\phi = 1,
\end{equation}
where $\Omega_\phi$ 
{is} the density parameter of the scalar field. We also define a `scalar fraction' $f_\phi$ as:
\begin{equation}
    f_\phi \equiv \frac{\Omega_\phi}{\Omega_\Lambda+\Omega_\phi},
\end{equation}
so that:
\begin{eqnarray}
    \Omega_\phi &=& f_\phi\left(1-\Omega_m\right),\nonumber\\
    \Omega_\Lambda &=& \left(1-f_\phi\right)\left(1-\Omega_m\right).
\end{eqnarray}
In this case we have:
\begin{equation}
    \beta = \frac{\frac{1}{2}\Omega_ma^{-3}+\Omega_\Lambda+\Omega_{\textrm{rc}}}{\sqrt{\Omega_{\textrm{rc}}\left(\Omega_ma^{-3}+\Omega_\Lambda+\Omega_{\textrm{rc}}\right)}},
\end{equation}
and the background expansion history is given as:
\begin{equation}
    E = \sqrt{\Omega_ma^{-3}+\Omega_\Lambda+\Omega_{\textrm{rc}}} - \sqrt{\Omega_{\textrm{rc}}}.
\end{equation}

\subsubsection{The cubic vector Galileon (\ac{cvG}) model}
\label{subsubsect:cvG}

The \ac{cvG} model extends the Galileon framework by introducing a massive, dynamical vector field, $A_\mu$, rather than a 
scalar 
{\ac{DoF}}. The theory 
{we study here is described by the} generalised Proca action \citep{Heisenberg2014}, constructed from the Faraday tensor $F_{\mu\nu} = \nabla_\mu A_\nu - \nabla_\nu A_\mu$ and the vector invariant $A^2 = A_\mu A^\mu$. The crucial nonlinear Vainshtein screening mechanism is triggered by the vector self-interaction described by $F^{\mu\nu}F_{\mu\nu}$:
\begin{equation}
      S_{\mathrm{cvG}} = \int\mathrm{d}^4x \sqrt{-g} \left[ \frac{M_{\mathrm{Pl}}^2}{2} R - \frac{1}{4} F_{\mu\nu}F^{\mu\nu} + \frac{1}{2} m^2 A^2 + \beta_3 A^2 \nabla_\mu A^\mu
      \right],
\end{equation}
where $\beta_3$ dictates the strength of the cubic coupling, {and $m$ is the mass of the vector field}. 

On cosmological scales, the temporal component of the vector field, {$A_0$}, drives the accelerated background expansion. During nonlinear structure formation, the longitudinal spatial mode of the vector field ($\phi$, where $A_\mu \sim \partial_\mu \phi$) strictly dominates the clustering dynamics, {and the transverse, or vector, mode is completely negligible \citep{Beckeretal2020b}}. Because this longitudinal mode mathematically mirrors the behaviour of a scalar Galileon, it can be seamlessly mapped into our (scalar) master Vainshtein equation. {Despite this similarity, the derivation of the N-body equations in the cvG model is considerably more complicated than the scalar Galileon case. We refer readers to \cite{Beckeretal2020b} for details.} For the cvG model, we have:
\begin{align}
  \label{eq:cvG_beta} \beta &= -\frac{2}{3} \left( \frac{\tilde{\beta}_3}{2\Omega_\phi} \right)^{1/3} \left[ \left( 2\frac{E'}{E} - 1 \right) E^2 + \Omega_\phi E^{-2} \right] + \frac{2}{3} \tilde{\beta}_3 \tag{36},\\
  \label{eq:cvG_alpha} \alpha &= \Omega_\phi \left( \frac{\tilde{\beta}_3}{2\Omega_\phi} \right)^{1/3} E^{-2} \tag{37},\\
  \label{eq:cvG_Rc} R_c^2 &= 2 \left( \frac{\tilde{\beta}_3}{2\Omega_\phi} \right)^{2/3} E^2 \tag{38},
\end{align}
where $\tilde{\beta}_3$ is a free dimensionless parameter of this model {related to $\beta_3$ as $\tilde{\beta}_3=\beta_3(c/H_0)^{-2}$}, $\Omega_\phi$ is again the density parameter of the scalar field defined in the same way as above {for \ac{nDGP}}. Note that the $\beta$ and $R_c^2$ parameters here differ from those in \cite{Beckeretal2020b}, {and are related to the latter through}:
\begin{eqnarray}
    \beta &=& \frac{2}{3}\beta_{\textrm{Becker}},\\
    R^2_c &=& 2R^2_{c,\textrm{Becker}}
\end{eqnarray}
This is 
{to ensure that the resulting equations have exactly} the same form of the master equation, Eq.~\eqref{eq:scalar_eom_code_units} and the N-body Poisson equation, Eq.~\eqref{eq:poisson_eqn_code_units}.

For numerical {considerations, it is useful to rewrite the N-body equations, Eq.~\eqref{eq:scalar_eom_background} and Eq.~\eqref{eq:poisson_eqn_code_units}, in the following revised form}:
\begin{eqnarray}
    \label{eq:scalar_eom_code_units_2}
    \nabla^2\varphi' + \frac{R_c'^2}{3\beta' a^4}\left[\left(\nabla^2\varphi'\right)^2-\nabla^i\nabla^j\varphi'\nabla_i\nabla_j\varphi'\right] = \frac{1}{\beta'}\Omega_m a\delta,
\end{eqnarray}
and: 
\begin{equation}\label{eq:poisson_eqn_code_units_2}
    \nabla^2\Phi = \frac{3}{2}\Omega_m a\delta + \alpha'\nabla^2\varphi',
\end{equation}
which are obtained through a field redefinition:
\begin{equation}
    \varphi' = \frac{\beta}{\beta_{\textrm{sDGP}}}\varphi,
\end{equation}
and we have also defined:
\begin{align}
    \beta' &= \beta_{\mathrm{sDGP}},\\ 
    \alpha' &= \alpha\frac{\beta_{\mathrm{sDGP}}}{\beta},\\
    R_c' &= R_c\frac{\beta_{\mathrm{sDGP}}}{\beta}.
\end{align}
{This is because the original $\beta$, defined in Eq.~\eqref{eq:cvG_beta}, is determined by $\tilde{\beta}_3$ which can vary wildly depending on the model parameter choice, while for simulations to produce accurate results stably we want $\beta$ to be more predictable: this is achieved by the rescaling.}

The background expansion history of this model is given by:
\begin{equation}\label{eq:cvG_background}
    E^2 = \frac{1}{2}\left(\Omega_ma^{-3}+\Omega_\Lambda\right) + \frac{1}{2}\sqrt{\left(\Omega_ma^{-3}+\Omega_\Lambda\right)^2+4\Omega_\phi},
\end{equation}
where, to be generic, we have again allowed for the presence of a cosmological constant with density parameter $\Omega_\Lambda$. Taking the derivative of the above expression yields the $E'/E$ term required to evaluate the $\beta$ parameter, {Eq.~\eqref{eq:cvG_beta}. Observe that Eq.~\eqref{eq:cvG_background} does not depend on the parameter $\tilde{\beta}_3$, which means that all cvG models have the same background expansion history specified by $\Omega_m$ and $\Omega_\Lambda$\footnote{{Note that throughout this work we assume a flat universe, for which $\Omega_m+\Omega_\Lambda+\Omega_\phi=1$.}}. In particular, we will see below that this is identical to the background equation of the scalar Galileon model, highlighting again the similarity between the two classes of models. $\tilde{\beta}_3$ does determine $\beta$, $\alpha$ and $R_c$, suggesting that these models do indeed differ at the perturbation level; nevertheless, we will see below that as $\tilde{\beta}_3\rightarrow0$ the cvG model reduces to the scalar Galileon.}

{The cvG model has been relatively less well-studied numerically. For some previous simulation works, see, e.g., \cite{Beckeretal2020b,Beckeretal2020}.}

\subsubsection{The cubic scalar Galileon (\ac{csG}) model}
\label{subsubsect:cvG}

The \ac{csG} model \citep{Deffayetetal2009b} is the quintessential Vainshtein screening model. It is built from a scalar field Lagrangian that inherently respects the Galilean shift symmetry $\partial_\mu \phi \to \partial_\mu \phi + c_\mu$, in flat spacetime, guaranteeing that the resulting equations of motion remain strictly second-order to {avoid} Ostrogradsky ghosts. The relevant gravitational and scalar action is given by:
\begin{equation}
S_{\mathrm{csG}} = \int\mathrm{d}^4x \sqrt{-g} \left[ \frac{M_{\mathrm{Pl}}^2}{2} R + c_2 X + \frac{c_3}{M^3} X \Box \phi
\right],
\end{equation}
in which $X \equiv -\frac{1}{2}\nabla_\mu \phi \nabla^\mu \phi$ ({this is the convention of \citealt{Barreira:2013}, which differs from the one used here, but the physical content remains identical}) is the canonical kinetic term, $M$ is the characteristic mass scale of the theory, and $c_2, c_3$ are dimensionless constants. The non-linear derivative coupling term, $X \Box \phi$, acts as the 
physical origin of the Vainshtein mechanism, 
suppressing the fifth force deep within {massive} dark matter halos while {enhancing} structure {formation} on quasi-linear scales. 

The csG model is effectively the 
{$\tilde{\beta}_3\rightarrow0$} limit of the cvG model, and its parameters map to {the master equations} as:
\begin{align}
  \label{eq:csG_beta} \beta &= \frac{2}{3} \frac{1}{\sqrt{6\Omega_\phi}} \left[ \left( 2\frac{E'}{E} - 1 \right) E^2 + \Omega_\phi E^{-2} \right],\\
  \label{eq:csG_alpha} \alpha &= -\frac{1}{6} \sqrt{6\Omega_\phi} E^{-2},\\
  \label{eq:csG_Rc} R_c^2 &= \frac{1}{3\Omega_\phi} E^2.
\end{align}
Note that $\beta$ above is the $\beta_2$, and $R_c^2$ above is $\beta_2/\beta_1$, in \cite{Barreira:2013}. To obtain these expressions, {we have assumed that the field follows the ``tracker'' solution \citep[see, e.g.,][for how the parameters $c_2, c_3$ are fully determined by the cosmological parameters on the tracker, and therefore the theory essentially has no additional free parameters]{Barreiraetal2014}. The tracker solution is described by:
\begin{equation}
    H\dot{\phi} = \xi H_0^2,
\end{equation}
with $\xi = \sqrt{6\Omega_\phi}$ and $\phi$ is the background scalar field (with the overbar omitted as usual). In the convention of \cite{Barreira:2013}, $c_2=-1$ and $c_3 = 6^{-1}(6\Omega_\phi)^{-1/2}$.}

The background expansion history of the csG model is {given by}:
\begin{equation}\label{eq:csG_background}
    E^2 = \frac{1}{2}\left(\Omega_ma^{-3}+\Omega_\Lambda\right) + \frac{1}{2}\sqrt{\left(\Omega_ma^{-3}+\Omega_\Lambda\right)^2+4\Omega_\phi},
\end{equation}
which is identical to that of the cvG model, given in Eq.~\eqref{eq:cvG_background}.

As mentioned above, the csG model is essentially the cvG model in the limit of $\tilde{\beta}_3\rightarrow0$. As a sanity check, we need to show that Eqs.~\eqref{eq:cvG_beta} -- \eqref{eq:cvG_Rc} reduce to Eqs.~\eqref{eq:csG_beta} -- \eqref{eq:csG_Rc}. We first observe that, as $\tilde{\beta}_3\rightarrow0$, the last term of Eq.~\eqref{eq:cvG_beta} can be neglected because it becomes much smaller than the first term therein. For the two models to be identical, we need to require that they have the same $\alpha\nabla^2\varphi$, or equivalently $\alpha\varphi$, which ensures that their respective Poisson equations are identical. This can be achieved by a field redefinition so that:
\begin{equation}
    \varphi_B = \frac{\alpha_A}{\alpha_{\text{B}}}\varphi_A = -\sqrt{6\Omega_\phi}\left(\frac{\tilde{\beta}_3}{2\Omega_\phi}\right)^{1/3}\varphi_A,\nonumber
\end{equation}
where $A, B$ stand for cvG and csG respectively. Then we have:
\begin{align}
    \nabla^2\varphi_B &= -\sqrt{6\Omega_\phi}\left(\frac{\tilde{\beta}_3}{2\Omega_\phi}\right)^{1/3}\nabla^2\varphi_A,\nonumber\\
    \left(\nabla^2\varphi_B\right)^2 &= 6\Omega_\phi\left(\frac{\tilde{\beta}_3}{2\Omega_\phi}\right)^{2/3}\left(\nabla^2\varphi_A\right)^2.\nonumber
\end{align}
Substituting this into the model-B version of the scalar field \ac{EoM}, this requires: 
\begin{equation}
\begin{split}
    -\sqrt{6\Omega_\phi} \left( \frac{\tilde{\beta}_3}{2\Omega_\phi} \right)^{1/3} \nabla^2 \varphi_A 
    + \frac{R_{c,B}^2}{3\beta_B a^4} 6\Omega_\phi & \\
    \left( \frac{\tilde{\beta}_3}{2\Omega_\phi} \right)^{2/3} \left[ (\nabla^2 \varphi_A)^2 - (\nabla^{ij} \varphi_A)^2 \right] 
    &= \frac{1}{\beta_B} \Omega_m a \delta.\nonumber
\end{split}
\end{equation}
For this to be identical to the model-A \ac{EoM}, we must then have:
\begin{equation}
    \frac{\beta_A}{\beta_B} = -\sqrt{6\Omega_\phi}\left(\frac{\tilde{\beta}_3}{2\Omega_\phi}\right)^{1/3},\nonumber
\end{equation}
and:
\begin{equation}
    \frac{R^2_{c,A}}{R^2_{c,B}} = 6\Omega_\phi\left(\frac{\tilde{\beta}_3}{2\Omega_\phi}\right)^{2/3}.\nonumber
\end{equation}
It is straightforward to verify that both conditions are satisfied.

{Because of this equivalence in the $\tilde{\beta}_3\rightarrow0$ limit, csG can be considered as a special case of cvG. This is how we treat the two classes of models in {\EFTRAMSES}, which only implements cvG explicitly, and achieves csG if $\tilde{\beta}_3$ is set to be a tiny value, e.g., $10^{-6}$. Together with the inclusion of $f_\phi$ that allows the cvG model to have an additional cosmological constant, the actual cvG model in {\EFTRAMSES} has two free parameters, $f_\phi$ and $\tilde{\beta}_3$, making it more flexible yet still covering all the well-known cases.}

\subsubsection{Generic EFT parameterisation}
\label{subsubsect:parameterised_eft}

Rather than a specific fundamental Lagrangian, this approach utilises the phenomenological parameterisation of the \ac{EFTofDE} \citep{Gubitosi:2013,Bellini2014}. As derived in Section \ref{subsect:eftofde}, the background and perturbation dynamics are entirely dictated by the time-dependent $\alpha$-basis functions. {This includes also the nonlinear perturbation dynamics, barring situations corresponding to clustering dark energy (see discussion above).} This makes it an incredibly flexible, model-agnostic umbrella for testing generic deviations from \ac{GR} strictly using observational data. 

By converting the master equation into the $\alpha$-basis, we can map the generic EFT parameters to our solver as:
\begin{align}
  \label{eq:eft_beta} \beta &= - \frac{2}{3 (\alpha_{\text{M}} - \alpha_{\text{B}})} \bigg[ \frac{3}{2}\Omega_{\text{DE}}(a)[1+w_{\text{DE}}(a)]\nonumber\\
  & ~~~~~~~~~~~~~~~~~~~~~~~~~~~~~+ (1+\alpha_{\text{B}})(\alpha_{\text{M}}-\alpha_{\text{B}})-\frac{E'}{E}\alpha_{\text{B}} - \alpha_{\text{B}}' \bigg],\\
  \label{eq:eft_alpha} \alpha &= \alpha_{\text{B}} - \alpha_{\text{M}},\\
  \label{eq:eft_Rc} R_c^2 &= \frac{\alpha_{\text{M}} - 2\alpha_{\text{B}}}{E^2 (\alpha_{\text{M}} - \alpha_{\text{B}})},
\end{align}
where $\beta$, $\alpha$ and $R_c^2$ are obtained using the methodology in Section \ref{subsect:code_units} following from Eq.~\eqref{eq:nbody_eqn_alpha} which governs the cross over terms between the N-body equation. 

To {calculate} these parameters dynamically throughout the simulation, the background expansion history must be explicitly defined. For our specific EFT implementation, we adopt the \ac{CPL} parameterisation \citep{ChevallierPolarski2001} for dynamical dark energy, where the equation of state evolves as:
\begin{equation}
    w_{\mathrm{DE}}(a) = w_0 + w_a(1-a),
\end{equation}
{where $w_0,w_a$ are constants.} Using this equation of state, we can find the physical dark energy density by solving the standard continuity equation. Defining the e-folding number as $N = \ln(a)$, the continuity equation for dark energy is given by:
\begin{equation}
    \frac{\mathrm{d}}{\mathrm{d}N} \ln\left(\frac{\rho_{\mathrm{DE}}}{\rho_{\mathrm{DE}0}}\right) = -3(1 + w_{\mathrm{DE}}(a)) = -3(1 + w_0 + w_a) + 3w_a e^N.\nonumber
\end{equation}
Integrating this expression from the present day ($N=0$) to a given scale factor $a$ yields:
\begin{equation}
    \rho_{\mathrm{DE}}(a) = \rho_{\mathrm{DE}0} \exp(-3w_a) a^{-3(1+w_0+w_a)} \exp(3w_a a).
\end{equation}
The corresponding dark energy density fraction and its evolution are governed by:
\begin{eqnarray}\label{eq:Omega_DE_generic}
    \Omega_{\mathrm{DE}}(a) = 1 - \Omega_{m}a^{-3} E^{-2},
\end{eqnarray}
where the normalised Hubble expansion rate derivative is given by:
\begin{equation}\label{eq:EpoE_generic}
     \frac{E'}{E} = -\frac{3}{2} \left[ 1 + w_{\mathrm{DE}}(a)\Omega_{\mathrm{DE}}(a) \right].
\end{equation} 
To close the system, the derivatives of the dark energy equation of state and density fraction with respect to N are evaluated as:
\begin{align}
    w_{\mathrm{DE}}'(a)& = -w_a a,\\ \Omega_{\mathrm{DE}}'(a) &= \left[1- \Omega_{\mathrm{DE}}(a)\right] \left[3+2\frac{E'}{E}\right].
\end{align}

Finally, for a {concrete example of the} EFT model to be simulated in this work, we assume the running Planck mass remains unmodified ($\alpha_{\text{M}} = 0$), while the braiding parameter is {assumed to be} proportional to the dark energy density {with a proportionality coefficient}, $c_{\text{B}}$:
\begin{equation}
   \alpha_{\text{B}} = c_{\text{B}} \Omega_{\mathrm{DE}}(a), \quad \quad \alpha_{\text{B}}' = c_{\text{B}} \Omega_{\mathrm{DE}}'(a).
\end{equation}
By feeding these background relations directly into the $\beta$, $\alpha$, and $R_c^2$ expressions, our pipeline can accurately simulate the fully non-linear spatial clustering of the \ac{EFT}.

\subsubsection{The generalised covariant cubic Galileon (\ac{GCCG}) model}
\label{subsubsect:gccg}

The \ac{GCCG} model \citep[see, e.g.,][]{Fruscianteetal2020,Ataydeetal2024} is {an} 
extension of the standard covariant Galileon. Traditional Galileon models often struggle with \ac{ISW} constraints because their background evolution is rigidly locked to the scalar field dynamics. To resolve this, the \ac{GCCG} {model} replaces the standard linear terms in the k-essence and kinetic braiding Lagrangians with arbitrary power laws of the kinetic energy. The defining gravitational and scalar action is given by:
\begin{equation}
\begin{split}
    S_{\mathrm{GCCG}} = \int\mathrm{d}^4x \sqrt{-g} \bigg[ &\frac{M_{\mathrm{Pl}}^2}{2} R + c_2 M^{4(1-p_2)} (-X)^{p_2} \\
    &+ c_3 M^{1-4p_3} (-X)^{p_3} \Box \phi 
    \bigg],
\end{split}
\end{equation}
where 
$M$ is the characteristic mass scale of the theory, and the exponents $c_2,c_3$ and $p_2,p_3$ are free dimensionless constants.

{We can follow the same procedure as before and derive the expressions for $\beta, \alpha$ and $R_c$ for the GCCG model. However, since we have already introduced the implementation of the all-powerful \ac{EFTofDE} parameterisation in the previous subsubsection, we can make use of the fact that} the GCCG model can be treated as a special case of the \ac{EFTofDE}. {With this in mind, we consider that other models can also be treated in the same way, and thus \ac{GCCG} will be the last specific \ac{MG} model to be considered explicitly here.}

The background expansion history {(on tracker)} is governed by the following algebraic equation 
\citep[e.g.,][]{Ataydeetal2024}:
\begin{equation}\label{eq:gccg_background}
    E^{s+2}- \Omega_{m}a^{-3}E^s - \Omega_{\textrm{DE}} = 0,
\end{equation}
and $\alpha_{\textrm{B}}$ is given by:
\begin{equation}\label{eq:gccg_alpha}
    \alpha_{\textrm{B}} = -s q \Omega_{\textrm{DE}}E^{-(s+2)} = \alpha,
\end{equation}
where $s,q$ are dimensionless model parameters with $s=2,q=1/2$ being the special case of cubic Galileon (on tracker).

{With $E$ solved for using Eq.~\eqref{eq:gccg_background}, $\Omega_{\text{DE}}(a)$ can now be computed with Eq.~\eqref{eq:Omega_DE_generic}. Meanwhile, $E'$ can be obtained by taking the derivative of Eq.~\eqref{eq:gccg_background} with respect to $N=\ln a$ and Eq.~\eqref{eq:EpoE_generic} then gives $w_{\text{DE}}(a)$. We thus can obtain $\beta$ and $R_c$ from Eqs.~\eqref{eq:eft_beta} and \eqref{eq:eft_Rc}, respectively. This completes all the necessary ingredients to simulate the \ac{GCCG} model based on the master equation. A similar procedure can actually be used for the other models above, such as \ac{cvG}.}


\section{The {\EFTRAMSES} code and tests}
\label{sect:eft-ramses}

In this section, we outline the computational methodology developed to simulate non-linear structure formation within the \ac{EFTofDE} framework. We begin by reviewing the underlying architecture and multigrid relaxation techniques of the \ECOSMOG{} N-body code. Building upon this foundation, we introduce our novel extension, \EFTRAMSES, which transitions the code from relying on model-specific solvers to a generalised, versatile computational engine. We then detail the systematic transformation of our physical field equations into dimensionless code units.
\subsection{\ECOSMOG{}}

\ECOSMOG{} \citep[Efficient COde for Simulating MOdified Gravity,][]{Lietal2012} is {the first efficiently \textsc{mpi} parallelised} N-body simulation code, 
designed to study nonlinear structure formation in \ac{MG} and dynamical \ac{DE} scenarios, by overcoming the previous computational limitations that were present. 

Before its development, simulations for \ac{MG} models like the \ac{DGP} model and {$f(R)$ gravity} required standard fixed-grid particle mesh codes \citep{KhouryWyman2009,Oyaizuetal2009,Chan:2009ew}. These early computational efforts were 
limited because {the use of a uniform grid resolution made it impossible to give the needed high force resolution in dense regions while keeping the overall computing time under control. Some of these codes} 
employ coarse spherical top-hat approximations that lost precision 
{on small scales. Early attempts to go beyond fixed grids include implementations of various \ac{MG} models in the \textsc{mlapm} \citep{Knebe:2001av} code \citep[e.g.,][]{Li:2009sy,Zhao:2010az,li:2010jl,Li:2010re,Zhao:2010qy,li:2011al,Brax:2011ja,Davis:2011pj}, which uses \ac{AMR} to achieve high force resolution in dense regions. However, this is still a serial code which ran on single cores, and was therefore unsuitable for modern large, high-resolution, simulations.} 

To overcome these difficulties, \ECOSMOG{} was built as a comprehensive extension of the publicly available {\RAMSES} code \citep {Teyssier2002} which contains efficient parallelisation. \ECOSMOG{} solves these bottlenecks by inheriting robust \ac{AMR} capabilities. This \ac{AMR} framework allows the code to dynamically increase spatial resolution {where needed. It is also} essential for accurately resolving extra scalar fields and \ac{MG} forces precisely where they are most nonlinear, without the high computational cost of a universally fine grid. The core challenge in simulating \ac{MG} scenarios is solving the highly-nonlinear equations that govern the extra dynamical \ac{DoF}. \ECOSMOG{} solves this by solving these scalar field equations on the mesh using {the} Newton-Gauss-Seidel relaxation method. A typical consequence of single-grid relaxation for non-linear problems is the slow convergence, this is handled by employing a multigrid relaxation technique arranged in V-cycles, applying it across the regular domain grid and all adaptive refinement levels. {This strategy has been used in other \ac{MG} codes such as \textsc{mg-gadget} \citep{Puchwein:2013lza}, \textsc{isis} \citep{Llinares:2013jza}, \textsc{mg-arepo} \citep{Arnold:2019vpg,Hernandez-Aguayo:2020kgq} and \textsc{mg-glam} \citep{Ruan:2021wup,Hernandez-Aguayo:2021kuh}.}

Initially, \ECOSMOG{} was successfully deployed to simulate $f(R)$ gravity, establishing it as a reliable tool for modelling theories governed by the chameleon screening mechanism \citep{Lietal2012,Brax:2012JCAP...10..002B,Brax:2013mua}. Following 
{this, it was} upgraded (known as \ECOSMOG-\textsc{V}) to accommodate theories that 
{feature} the Vainshtein screening mechanism, such as the \ac{DGP} model \citep{Li:2013nua}. Unlike chameleon models, Vainshtein screening models contains highly non-linear derivative self couplings that are notoriously difficult to resolve. To ensure numerical stability and prevent the Newton-Gauss-Seidel relaxation process from converging to imaginary square roots, the operator-splitting method introduced in \cite{Chan:2009ew} was {extended and re-interpreted \citep{Li:2013nua}. These developments, along with {\ECOSMOG}'s efficient parallelisation,}  allowed the \ac{AMR} {simulations} to precisely capture the subtle screening effect on small, non-linear scales without losing force resolution. 

The stability of this new \ECOSMOG-\textsc{V} framework opened an avenue for simulations of even more complex scenarios. The first cosmological simulations of the cubic Galileon gravity were achieved \citep{Barreira:2013}, demonstrating that the Vainshtein mechanism successfully suppressed the fifth force in haloes. The code was subsequently extended to simulate the quartic Galileon model \citep{Li:2013tda}. In this particular model, the field equations involve complex products of second-order derivatives, the numerical algorithm was altered to treat the field equation as a cubic algebraic equation for the field's Laplacian. 
Simulating modified gravity is computationally expensive. 
\cite{Barreiraetal2015c} {showed} that in highly-refined, high-density regions, the Vainshtein mechanism is so efficient that the fifth force becomes virtually negligible. By 
truncating the multigrid iteration for the scalar field on the finer refinement levels, and instead simply interpolating the solutions from coarser {levels}, the performance of the code was {improved} by up to a factor of $\simeq10$. This truncation sacrifices almost no accuracy, having less than a $1\%$ impact on the matter power spectrum, 
because the numerical error is induced only on a force that is already vanishingly small.

\subsection{\EFTRAMSES{}}

While \ECOSMOG{} has proven to be a vital computational tool, its earlier works were 
model-specific. Implementing new gravity theories such as transitioning from $f(R)$ gravity to the various Galileon{-type} models, required rigorous derivations and hard-coding entirely new, complex, \ac{EoM} into the solver. Given the wide family of proposed \ac{DE} and \ac{MG} models, testing {individual theories} on a case-by-case basis using custom N-body simulations is highly inefficient. 

To tackle this issue, we introduce \EFTRAMSES, a comprehensive extension of the \ECOSMOG{} code that incorporates the \ac{EFTofDE} framework as discussed in Section \ref{subsect:eftofde}. Rather than relying on specific theoretical constructs, \EFTRAMSES{} integrates the generalised mathematical framework of the $\alpha$-basis \citep{Bellini2014} parameters directly into the existing \ECOSMOG{} infrastructure. In this manner, the extra scalar \ac{DoF} is strictly defined in terms of the \ac{EFT} parameters, giving \EFTRAMSES{} the capability to solve the generalised, nonlinear \ac{EoM} for virtually any single-field {Vainshtein-screening \ac{MG}} theory 
{with minimal and user-friendly modifications} to the underlying code, {and without touching its core part}. Crucially, {\EFTRAMSES} 
{inherits} all the numerical advantages historically developed for \ECOSMOG-\textsc{V} as mentioned previously. 

Ultimately, \EFTRAMSES{} transitions the {\ECOSMOG} code from a collection of specialised solvers into {a unified framework for luminal cubic Horndeski theories } for exploring the non-linear regime structure formation. 
This updated version 
{covers} distinct theories {described in Section \ref{sect:individual_models}}, such as the \ac{DGP} model, various Galileon models, {as well as} the broader {and phenomenological} \ac{EFT} parameterisation, which {all} mathematically map to a single, generalised master Vainshtein equation{, Eq.~\eqref{eq:scalar_eom_code_units}.} 

%

\subsection{Setup of test simulations}
\label{subsect:test_sims}

To validate the numerical implementation of {\EFTRAMSES}, we performed cosmological N-body simulations for several Vainshtein-type models, and compared the results with previous or other simulations. All the test simulations were carried out in a periodic cubic box with a comoving side length of $L_{\text{box}} = 1024 \, h^{-1}\text{Mpc}$, using $N_{\textrm{p}}=1024^3$ dark matter particles. {We did this for the \ac{nDGP}, \ac{csG}, \ac{GCCG} models, as well as a parameterised \ac{EFT} theory.}

To ensure a robust, one-to-one comparison between the standard $\Lambda$CDM baseline and the various modified gravity scenarios, all simulation runs were seeded using the exact same \ac{IC}. By using the identical initial density fluctuations generated at a starting redshift of $z_{\text{ini}} = 49$, and computing the \ac{MG}-induced enhancements in matter clustering, we reduce the statistical noise {induced} by {sample} variance. This allows us to maximally isolate the differences driven purely by the modified gravitational forces and the different background expansion histories. Since our objective here is to compare with existing simulations where possible, we have used existing \ac{IC}s, and as a result the \ac{IC}s are different for the different \ac{MG} models. The complete set of cosmological and model-specific parameters is summarised in Table~\ref{tab:mg_parameters}.

\begin{table*}
    \centering
    \renewcommand{\arraystretch}{1.3} 
    \begin{tabular}{lccccccccc}
        \hline
        \hline
        {Model} & $L_{\text{box}}$ ($h^{-1}\mathrm{Mpc}$) & $N_{\text{p}}$ & $\Omega_{\text{m}}$ & $h$ & \texttt{param\_bg\_1} & \texttt{param\_bg\_2} & \texttt{param\_pt\_1} & \ac{IC} Code & Variants \\
        \hline
        \ac{nDGP} & $1024$ & $1024^3$ & $0.281$ & $0.697$ & --- & --- & $\Omega_{\text{rc}}=0.250$ & \textsc{MPGRAFIC} & Full/$\Lambda$CDM\\
        \ac{csG}   & $1024$ & $1024^3$ & $0.314$ & $0.674$ & $f_\phi=1.000$ & --- & --- & \textsc{2LPTic} & Full/QCDM/$\Lambda$CDM \\
        \ac{GCCG} & $1024$ & $1024^3$ & $0.256$ & $0.738$ & $s=0.649$ & --- & $q=1.061$ & \textsc{2LPTic} & Full/QCDM/$\Lambda$CDM \\
        \ac{EFT}N & $1024$ & $1024^3$ & $0.300$ & $0.700$ & $w_0=-1.000$ & $w_a=-0.200$ & $c_{\text{B}}=-1.000$ & \textsc{2LPTic} & Full/QCDM/$\Lambda$CDM \\
        \ac{EFT}P & $1024$ & $1024^3$ & $0.300$ & $0.700$ & $w_0=-1.000$ & $w_a=-0.200$ & $c_{\text{B}}=1.000$ & \textsc{2LPTic} & Full/QCDM/$\Lambda$CDM \\
        \hline
        \hline
    \end{tabular}
    \caption{Parameters {and specs} for all simulated models, {where the \ac{MG} parameters are} mapped to the generalised EFT background (\texttt{param\_bg\_i}) and perturbation (\texttt{param\_pt\_i}) functions. {Note that in some cases an \ac{MG} parameter affects both the expansion history and perturbation growth, in which case we consider it as a \texttt{param\_bg}. For \ac{nDGP}, we have considered a $\Lambda$CDM background, so the only \ac{MG} parameter is $\Omega_{\text{rc}}$, cf.~Eqs.~(\ref{eq:nDGP_beta}, \ref{eq:nDGP_Rc}). For \ac{csG}, we assume there is no cosmological constant, so $f_\phi=1$, i.e., $\Omega_\Lambda=0$ in Eq.~\eqref{eq:csG_background}; the model has otherwise no free parameters. For \ac{GCCG}, $s$ is a \texttt{param\_bg} because it determines the background expansion, cf.~Eq.~\eqref{eq:gccg_background}, while $q$ is a \texttt{param\_pt} as it determines $\alpha_{\text{B}}$, cf.~Eq.~\eqref{eq:gccg_alpha}. Finally, for \ac{EFT}, it is clear that $w_0,w_a$ are \texttt{param\_bg} while $c_{\text{B}}$ is a \texttt{param\_pt}. All simulations other than \ac{nDGP} used \ac{IC}s generated using the \textsc{2LPTic} code \citep{Crocce:2006ve}, while the \ac{IC} for the \ac{nDGP} simulations were generated using \textsc{MPGRAFIC} \citep{Prunet:2008fv}. Only \ac{nDGP} has a $\Lambda$CDM background, which is why for it we only have the `Full' simulation and its `$\Lambda$CDM' counterpart, while for all other \ac{MG} models we also have a `QCDM' counterpart. {The `counterpart' of an \ac{MG} model is where the cosmological parameters and \ac{IC}s are the same, but the fifth force is artificially set to $0$ in the simulation.} All models have only one simulation realisation, apart from \ac{csG} which has two realisations using \ac{IC}s with reverted phases to reduce sample variance \citep{Angulo:2016hjd}. All simulations assume a flat universe, and $h\equiv H_0/(100\,\mathrm{km}\,\mathrm{s}^{-1}\mathrm{Mpc}^{-1})$.}}
    \label{tab:mg_parameters}
\end{table*}

\subsubsection{The nDGP Model}
\label{subsubsect:nDGP_tests}

\begin{figure}
    \centering
    
    \includegraphics[width=\columnwidth]{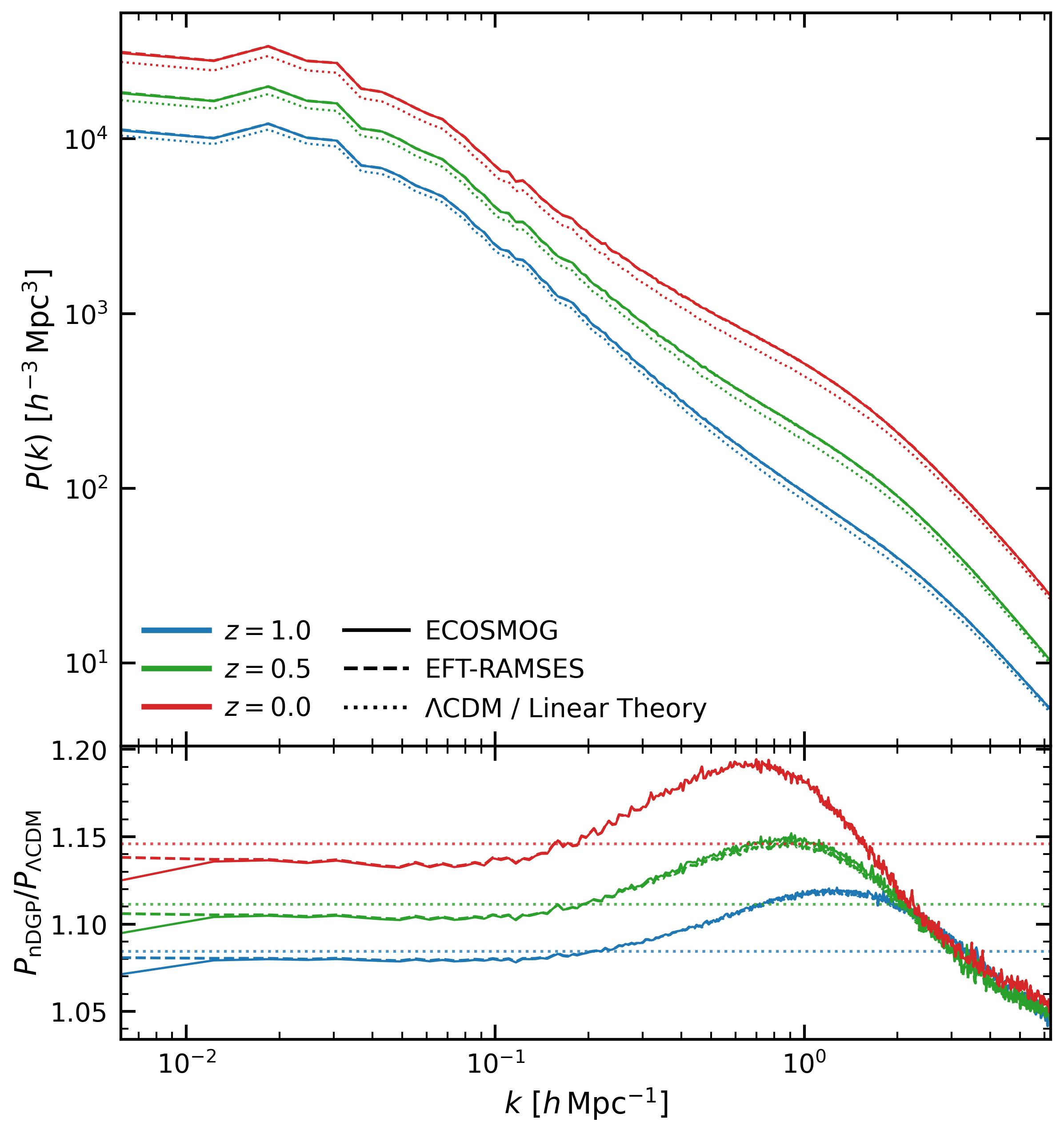}
    \caption{{The non-linear matter power spectrum of the nDGP model across three redshifts ($z = 1.0, 0.5, 0.0$). \textit{Top Panel:} The absolute power spectrum $P(k)$ in physical units, comparing the \ac{MG} runs to the $\Lambda$CDM baseline (dotted lines). \textit{Bottom Panel:} The fractional enhancement of the nDGP model relative to $\Lambda$CDM. In both panels, the dashed lines represent the predictions by \textsc{eft-ramses}, while the solid lines represent the results of the old \textsc{ecosmog} code. The two codes are in near-perfect agreement. The analytic linear theory enhancements are shown as dotted horizontal lines in the bottom panel. The distinct turnover at high $k$ illustrates the effect of the Vainshtein screening mechanism within dense halos.}}
    \label{fig:ndgp_redshift1}
\end{figure}

Figure \ref{fig:ndgp_redshift1} presents the non-linear matter power spectrum (top panel) and its enhancement relative to the $\Lambda$CDM baseline (bottom panel), {for the \ac{nDGP} model, at three distinct redshifts ($z=1.0, 0.5$, and $0$). We compare \EFTRAMSES{} (dashed lines) against an older, established version of the \ECOSMOG-\textsc{V} (solid lines) code 
\ECOSMOG-\textsc{V}} differs from \EFTRAMSES{} mainly in a different field redefinition and how the background quantities (i.e., the coefficients in the Master and modified Poisson equations) are calculated. 
Generally, {we find that,} across the entire range from $k \approx 0.005~h\mathrm{Mpc}^{-1}$ to $k \approx 6.0~h\mathrm{Mpc}^{-1}$, the predictions of {both codes agree to within sub-percent precision ($\le 0.5\%$).}

The bottom panel of Figure \ref{fig:ndgp_redshift1} shows the power spectrum enhancement of the \ac{nDGP} model relative to the baseline $\Lambda$CDM. The 
impact of the fifth force is captured across three distinct spatial regimes. On large, linear, scales ($k \lesssim 0.1~h\mathrm{Mpc}^{-1}$), the 
linear theory predicts a scale-independent enhancement of about $8.4\%$, $11.1\%$, and $14.6\%$ for redshifts $z=1.0,0.5$ and $0.0$, respectively marked by the dotted horizontal lines. 
{The mismatch of the longest-wavelength (lowest-$k$) mode is due to some version difference that affects the new and old $\Lambda$CDM runs, where \EFTRAMSES{} shows a better behaviour.}

While {{both} N-body {codes}} broadly track {the linear-theory} predictions, a slight 
suppression {of power} is clearly visible at the {smallest} resolved $k$ modes ($k \lesssim 0.05~h\mathrm{Mpc}^{-1}$). For example, at $z=0$, the simulated enhancement 
{is about $1\%$ lower than the linear-theory prediction, and the difference is progressively smaller at higher redshifts}. 
{The same suppression was also found in \textsc{ecosmog-eft} \citep{Ganjoo:2026ugf} which is based also on \ECOSMOG{} and \textsc{mg-glam} \citep{Hernandez-Aguayo:2021kuh}, as well as \AREPO{} \citep{Hernandez-Aguayo2021}, yet another independent code that uses a similar grid-based algorithm (see below).} We dedicate a detailed discussion in Appendix \ref{sec:appendix1}, {which suggests that} this 
is an inherent mode-coupling effect driven by the non-linearities of the Vainshtein equation and showing {a comparison of this model between \EFTRAMSES{} and \AREPO{} codes on large scales}. Specifically, diagnostic simulations confirm that artificially suppressing this non-linear term {--- either directly or indirectly through a coarser resolution --- greatly} restores the analytic linear-theory prediction on large scales. 

In the quasi-linear regime ($0.1 \lesssim k \lesssim 1.0~h\mathrm{Mpc}^{-1}$), the unsuppressed fifth force accelerates the clustering of dark matter, driving an increase {in $P(k)$ enhancement} that peaks at $k\sim0.7~h\mathrm{Mpc}^{-1}$ 
({with a maximum value of} $\sim19.5\%$) at $z=0$, {moving to $k\sim1.2~h\mathrm{Mpc}^{-1}$ at $z=1$}. Finally, deep within the non-linear regime ($k \gtrsim 1.0~h\mathrm{Mpc}^{-1}$), the curves show a distinct turnover, decaying back toward unity. This {turnover} is 
{a} signature of the Vainshtein screening mechanism dynamically activating within high-density collapsed halos. Accurately capturing this high-$k$ screening turnover relies entirely on the code's \ac{AMR} architecture, which dynamically deploys finer meshes to resolve the steep spatial gradients of the scalar field. 

{These full \ac{AMR}} simulations 
were initialised using {what we call} ``educated guess'' for the scalar field solver ($\phi_0 \propto \Phi$). As detailed in Appendix \ref{sec:appendix1}, diagnostic tests without refinement reveal that while this {educated} guess marginally reduces the large-scale mismatch {with linear theory when using} coarse simulation grids, it is not enough to completely eliminate it. More importantly, we observe {here} that this mild improvement is {lost} once full \ac{AMR} is activated. 

\subsubsection{The \ac{csG} Model}
\label{subsubsect:cG_tests}

In this {subsubsection}, we {specialise} \EFTRAMSES{} to the \ac{csG} model. Unlike the \ac{nDGP} model, which {has} identical background expansion history as standard $\Lambda$CDM, the \ac{csG} model alters the expansion rate. To accurately disentangle the effect of this modified background from {that of} the fifth force, we also simulate the counterpart \ac{QCDM} model, a baseline that strictly assumes \ac{GR} {for the gravity force} but evolves under the specific expansion history of the \ac{csG} model.

\begin{figure}
    \centering
    \includegraphics[width=\columnwidth]{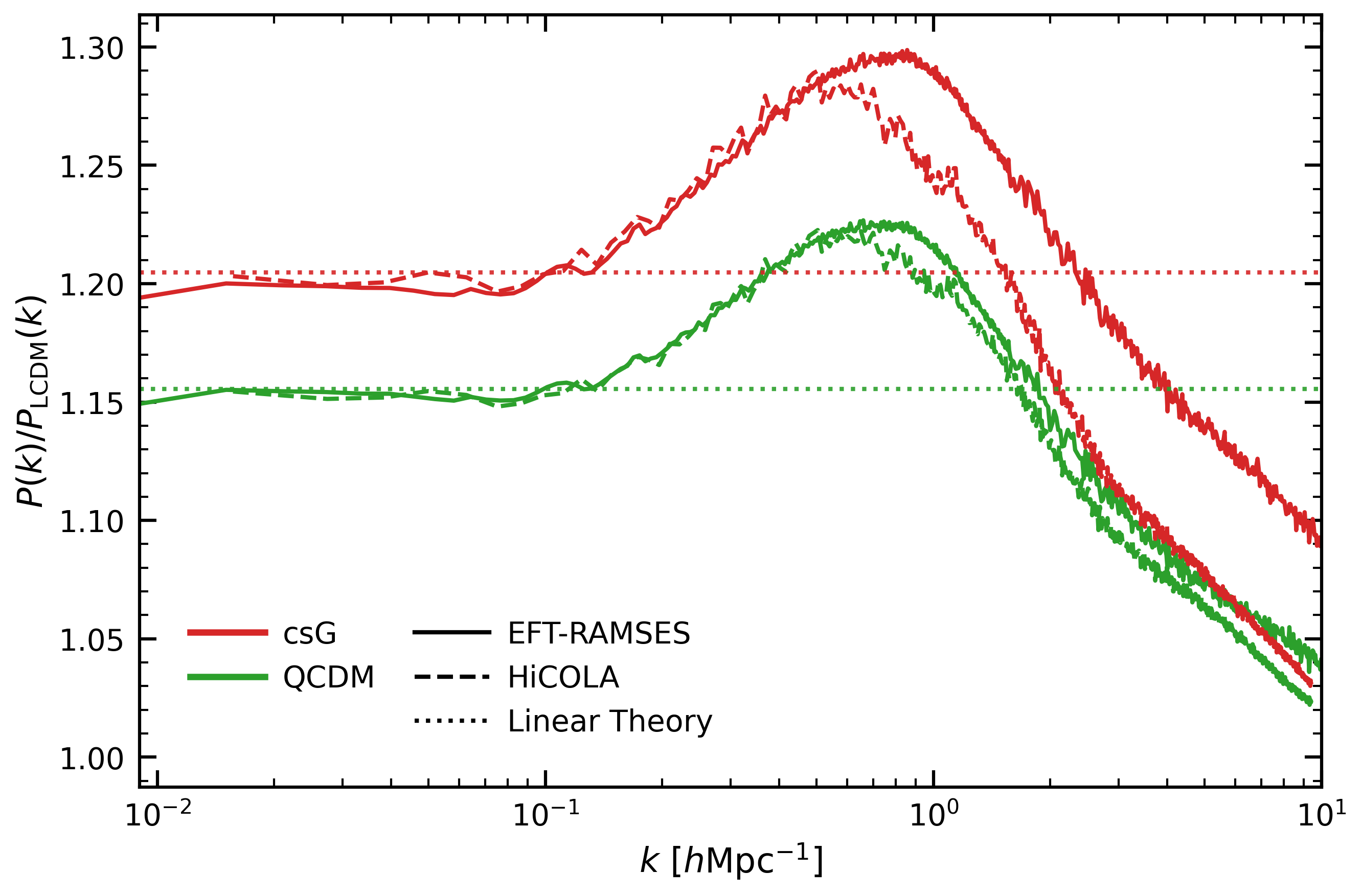}
    \caption{The fractional enhancement of the non-linear matter power spectrum at $z=0$ for the \ac{csG} (red) and \ac{QCDM} (orange) models. The solid lines denote the \EFTRAMSES{} simulations, which have been averaged over two {inverted}-phase realisations to suppress sample variance at low $k$. The dashed lines show the predictions from the \HICOLA{} code for comparison. The dotted horizontal lines indicate the linear-theory expectations based on the linear growth factors. The two N-body {codes} exhibit excellent agreement up to $k \approx 1.0~h\mathrm{Mpc}^{-1}$, successfully capturing the identical quasi-linear fifth-force bump and the onset of the Vainshtein screening turnover.}
    \label{fig:cg_qcdm_master}
\end{figure}

We compare our implementation against \HICOLA\footnote{\url{https://github.com/Hi-COLACode/Hi-COLA}}~\citep{Wright:2023}, an independent, approximate N-body simulation code specifically designed for fast evaluation of screening mechanisms. Figure~\ref{fig:cg_qcdm_master} presents the fractional enhancement of the matter power spectrum at $z=0$ for both the \ac{csG} {(red)} and \ac{QCDM} (orange) models relative to the $\Lambda$CDM baseline. To suppress {sample variance} and ensure an accurate comparison with {linear} theory, {we generated paired \ac{IC}s using the technique of \cite{Angulo:2016hjd}, and} the \EFTRAMSES{} power spectra were calculated by averaging the two simulations. 

On large, linear, scales ($k < 0.1~h\mathrm{Mpc}^{-1}$), the 
variance cancellation yields 
smooth curves that are in excellent agreement with the 
linear-theory predictions (dotted horizontal lines). For the cosmology studied here, the exact linear growth factors at $z=0$ are $D_{\mathrm{cG}} \approx 0.864$, $D_{\mathrm{QCDM}} \approx 0.846$, and $D_{\Lambda\mathrm{CDM}} \approx 0.787$. 

The \EFTRAMSES{} simulation {for \ac{QCDM} recovers its} large-scale limits {perfectly}, confirming that the distinct background expansion 
{is} correctly implemented in the {code. For \ac{csG}, however, we notice the same large-scale mismatch of the enhancement at $k\lesssim0.1~h\mathrm{Mpc}^{-1}$, though with a smaller amplitude than for \ac{nDGP}}.

{In} the quasi-linear regime ($0.1 \lesssim k \lesssim 1.0~h\mathrm{Mpc}^{-1}$), the fifth force drives an enhancement in {matter} clustering, peaking at roughly $30\%$ for the full \ac{csG} model and $22\%$ for the \ac{QCDM} model. {This shows that much of the difference between the \ac{csG} model and its $\Lambda$CDM counterpart is driven by the modified expansion history -- in fact, even the shape of the \ac{QCDM} enhancement looks similar to that of \ac{csG}, closely mimicking the Vainshtein screening effect.} In this regime, the \EFTRAMSES{} predictions (solid lines) show excellent agreement with the \HICOLA{} simulations (dashed lines), cross-validating the fifth force independent simulation pipelines. 

{In} the nonlinear regime  ($k \gtrsim 1.0~h\mathrm{Mpc}^{-1}$), both models exhibit {the} characteristic turnover due to the Vainshtein screening mechanism. While both codes capture the shape of this suppression, a slight difference between \EFTRAMSES{} and \HICOLA{} emerges at the {smallest} scales  ($k > 2.0~h\mathrm{Mpc}^{-1}$). {This is consistent with the result of \cite{Bose:2024qbw} which compared \ECOSMOG{} and \HICOLA, and is a result of the approximation used by \HICOLA{} to deal {with} the fifth force in this regime.} 

{We note that, for the full \ac{csG} model, \HICOLA{} does not suffer from the large-scale mismatch with the linear theory enhancement prediction, and this is likely precisely because it does not solve the master equation.}

\subsubsection{The \ac{GCCG} Model}
\label{subsubsect:GCCG_tests}

\begin{figure}
 \includegraphics[width=\columnwidth]{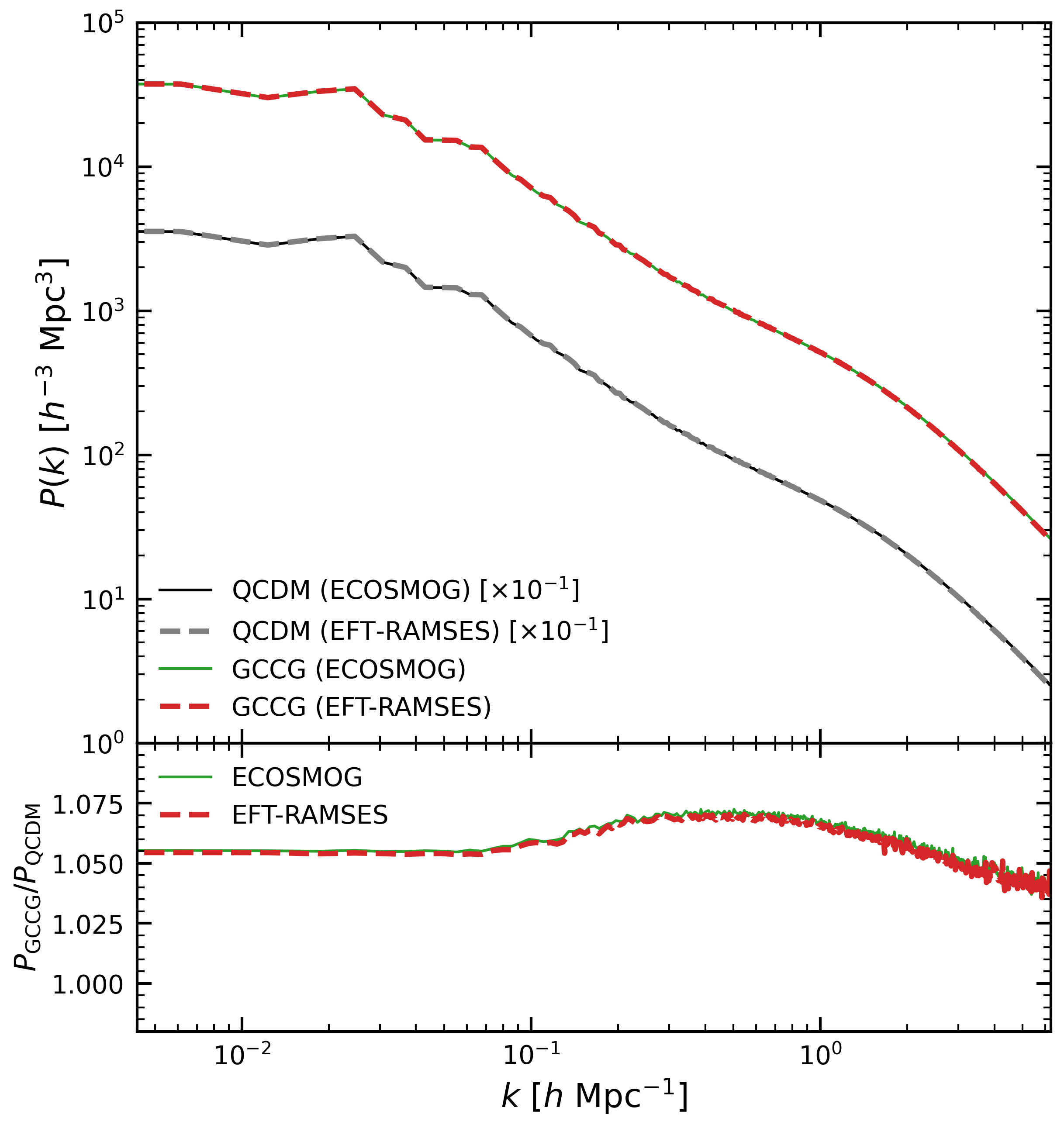}
 \caption{Comparison of the non-linear matter power spectra at $z=0$ for the QCDM and GCCG cosmologies. {The \ECOSMOG-\ac{QCDM} and \ECOSMOG-\ac{GCCG} data are preliminary results from work in preparation by \citet{Atayde:2026}}. \textit{Top panel}: The absolute matter power spectra, $P(k)$. To improve visual clarity and prevent overlapping, the QCDM curves have been artificially shifted downward by a decade ($\times 10^{-1}$). The outputs from {\ECOSMOG{}} (thin solid lines) are plotted beneath the outputs from \EFTRAMSES{} (thick dashed lines). The visual indistinguishability of the curves shows excellent code agreement. \textit{Bottom panel}: The physical enhancement of structure formation driven by the fifth force, represented by the ratio $P_{\text{GCCG}} / P_{\text{QCDM}}$.
 }
 \label{GCCGFigure}
\end{figure}

{Here we compared \EFTRAMSES{}} against {a modified version \citep{Atayde:2026} of \ECOSMOG{} to simulate the \ac{GCCG} model} \citep{Ataydeetal2024}. {\citet{Atayde:2026} is the first attempt of simulating \ac{GCCG} models: it differs from \EFTRAMSES{} in that the master equation takes a different form using an alternative field redefinition, and that the background cosmology has been pre-calculated numerically to produce a table that the code reads to compute background quantities by interpolation.} Figure \ref{GCCGFigure} shows this comparison at redshift $z=0$. 

The top panel shows the absolute matter power spectra for \ac{QCDM} and \ac{GCCG}{, where} the \ac{QCDM} curves have been artificially shifted downward by one decade ($\times 10^{-1}$) for visual clarity. Across the entire resolved {$k$} range, the {predictions by} \EFTRAMSES{} (thick dashed lines) are visually indistinguishable from the \cite{Atayde:2026} \ECOSMOG{} simulation results (thin solid lines), demonstrating excellent code agreement. The lower panel of Figure \ref{GCCGFigure} isolates the physical impact of the fifth force by plotting the fractional enhancement of structure formation, $P_{\text{GCCG}} / P_{\text{QCDM}}$, {again showing excellent agreement.} 

{Since simulation studies of the \ac{GCCG} model have not been presented elsewhere before, we refer the readers to \citet{Atayde:2026} for a more comprehensive analysis and presentation of its cosmological predictions.}


\subsubsection{The \ac{EFT} Parameterisation}
\label{subsubsect:eft_tests}

{Finally, since \EFTRAMSES{} is developed mainly for \ac{EFT}, in this subsubsection we show the results of some example simulations for the general Horndeski theory parameterised using \ac{EFT} functions (see Table \ref{tab:mg_parameters} for the parameters used). We have simulated two different \ac{EFT} models with the same $w_0$ and $w_a$, but different signs of $\alpha_\text{B}$, and refer to the model with $\alpha_{\text{B}}<0$ as \ac{EFTN}, while the one with $\alpha_{\text{B}}>0$ as \ac{EFTP}. \ac{EFTN} enhances while \ac{EFTP} suppresses structure formation relative to \ac{QCDM}.}

\begin{figure*}
    \centering
   
    \includegraphics[width=0.95\textwidth]{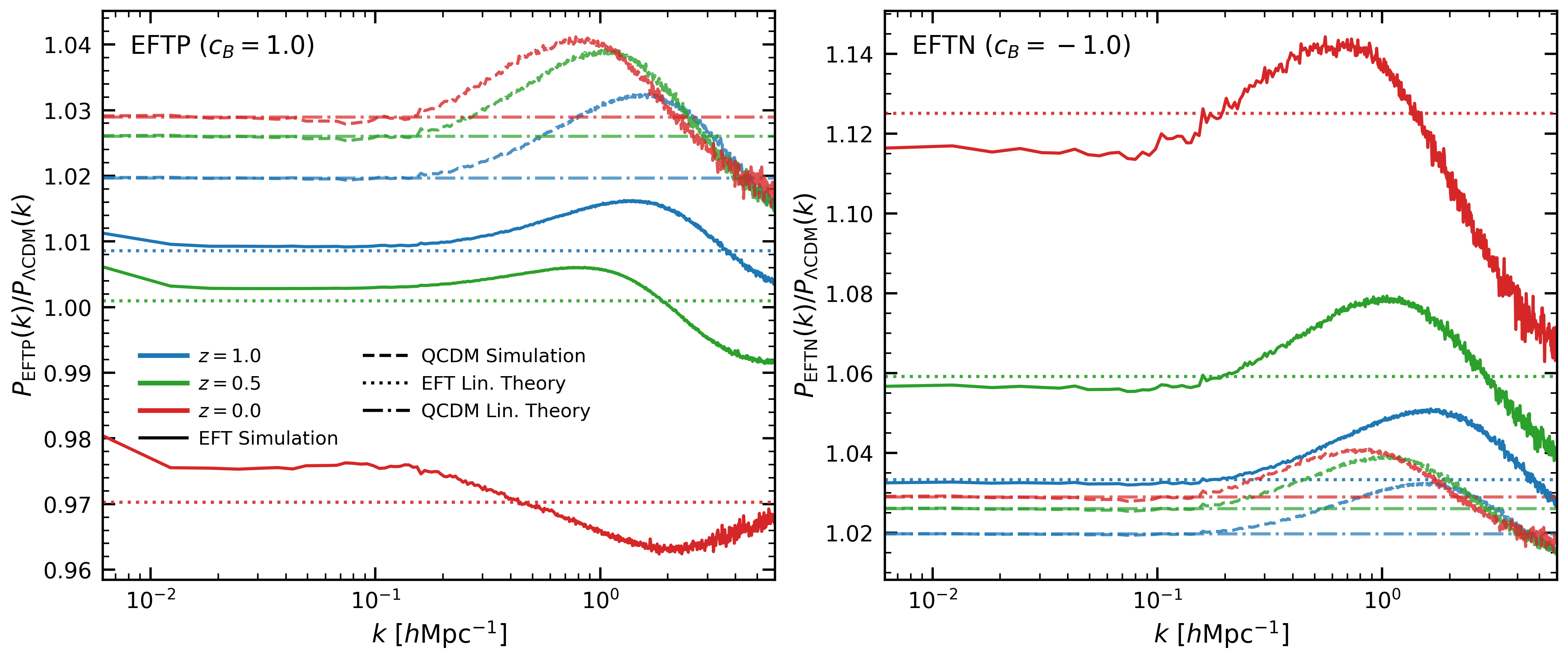}
    \caption{The fractional enhancement of the non-linear matter power spectrum for the \ac{EFT} and \ac{QCDM} models relative to the $\Lambda$CDM counterpart. Results are shown across three redshifts: $z=1.0$ (blue), $z=0.5$ (green), and $z=0.0$ (red). The left panel displays the \ac{EFTP} model ($c_{\text{B}} = 1.0$), while the right panel displays the \ac{EFTN} model ($c_{\text{B}} = -1.0$). Solid lines denote the full N-body simulations from \EFTRAMSES{}, and dashed lines represent the corresponding \ac{QCDM} simulations, which isolate the clustering effect driven purely by the modified background expansion. The dotted and dash-dotted horizontal lines indicate the linear-theory predictions for the \ac{EFT} and \ac{QCDM} models, respectively. 
    }
    \label{fig:eft_lcdm_enhancement}
\end{figure*}

\begin{figure*}
    \centering
    
    \includegraphics[width=0.95\textwidth]{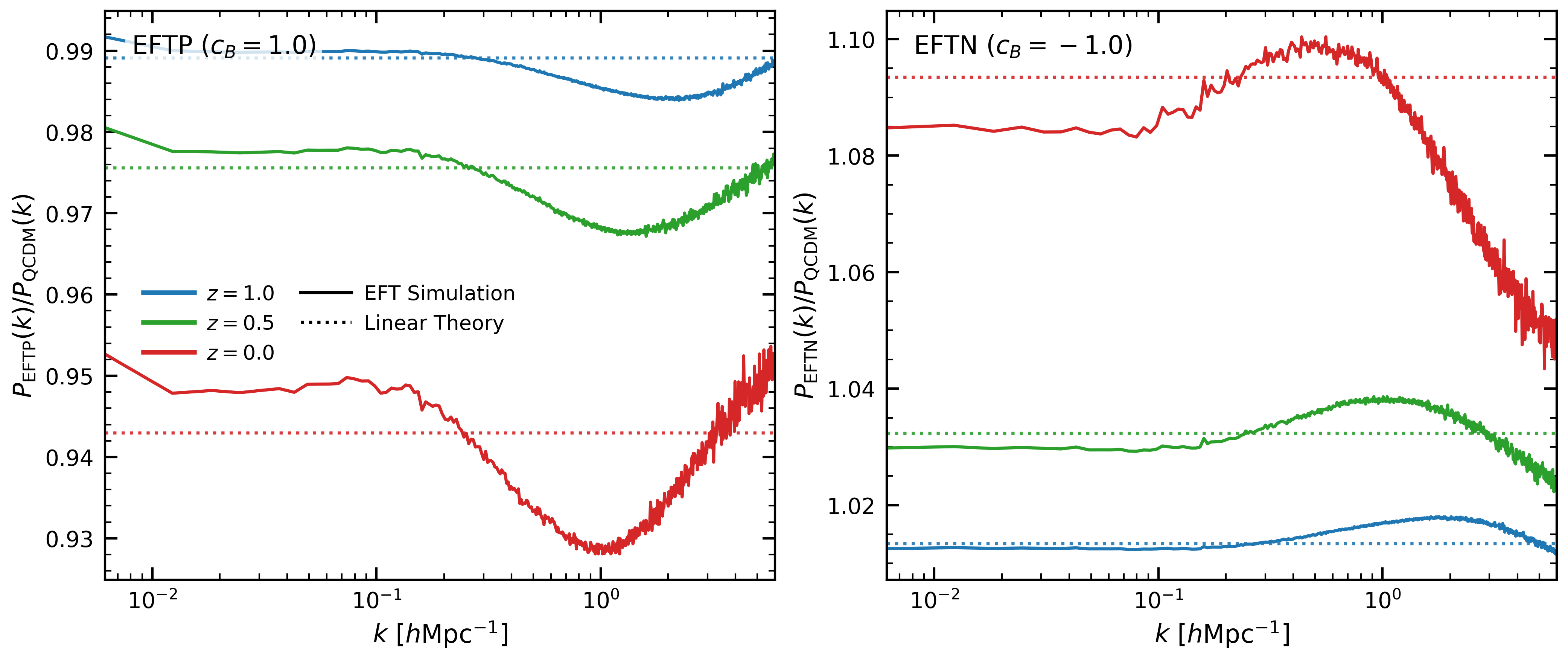}
    \caption{The fractional enhancement of the non-linear matter power spectrum for the \ac{EFT} model relative to the \ac{QCDM} counterpart. Results are shown across three redshifts: $z=1.0$ (blue), $z=0.5$ (green), and $z=0.0$ (red). The left panel displays the \ac{EFTP} model ($c_{\text{B}} = 1.0$), which suppresses structure formation relative to \ac{QCDM}. The right panel displays the \ac{EFTN} model ($c_{\text{B}} = -1.0$), which enhances structure formation. Solid lines denote the full N-body simulations from \EFTRAMSES{}, while dotted horizontal lines represent the corresponding linear-theory predictions. The overall strength of the fifth force grows progressively over cosmic time. 
    }
    \label{fig:eft_qcdm_baseline}
\end{figure*}

To study the effects of the \ac{MG} dynamics on structure formation, we present the non-linear matter power spectra from our \EFTRAMSES{} simulations of both models. Since the generic \ac{EFT} parameterisation inherently alters the background expansion of the Universe, it is instructive to compare the \ac{EFT} models against two distinct baselines. Figure \ref{fig:eft_lcdm_enhancement} presents the power spectra relative to the $\Lambda$CDM counterpart, which shows the combined impact of the modified background expansion (shown explicitly by the \ac{QCDM} curves) and the fifth force. To rigorously isolate the clustering effects driven purely by the \ac{MG} force, Figure \ref{fig:eft_qcdm_baseline} presents the fractional enhancement directly relative to the \ac{QCDM} counterpart, $P_{\mathrm{EFT}}(k) / P_{\mathrm{QCDM}}(k)$. In both figures, results are plotted across three redshifts: $z=1.0$ (blue), $z=0.5$ (green), and $z=0.0$ (red), with the left and right panels displaying the \ac{EFTP} and \ac{EFTN} models, respectively. The horizontal lines denote the corresponding linear-theory predictions, $(D_{\mathrm{EFT}} / D_{\mathrm{QCDM}})^2$. For \ac{EFTN}, these asymptotic baselines are 1.013 at $z=1.0$, 1.032 at $z=0.5$, and 1.093 at $z=0.0$. For \ac{EFTP}, the linear theory predicts a suppression of structure formation {with respect to \ac{QCDM}}, with $(D_{\mathrm{EFT}} / D_{\mathrm{QCDM}})^2$ of approximately $0.989$ at $z=1.0$, $0.976$ at $z=0.5$, and $0.943$ at $z=0.0$.


Several key physical features of the \ac{EFT} models are immediately apparent from this redshift evolution, particularly in Figure \ref{fig:eft_qcdm_baseline} where the fifth-force effect is isolated. Firstly, the overall strength of the fifth force grows progressively over cosmic time. For the \ac{EFTN} model, the clustering enhancement is small at $z=1.0$, peaking at less than $2\%$. By $z=0.5$, this enhancement reaches approximately $4\%$, and at the present day it nears $10\%$. Conversely, the \ac{EFTP} model exhibits a progressive suppression of structure formation relative to \ac{QCDM}. At $z=1.0$, the suppression is roughly $1.5\%$, which grows to over $3\%$ by $z=0.5$, and approximately $7\%$ at $z=0.0$ in the quasi-linear regime. Second, the large-scale mismatch with linear theory and the peaking and turnover due to Vainshtein screening are both qualitatively similar to---though quantitatively different from, as expected---the other \ac{MG} models discussed above. As such we do not {describe them in detail here}.

{We note that for \ac{EFTP}, the Vainshtein screening also suppresses the fifth force, making the predicted non-linear power spectrum approach \ac{QCDM} at small scales (exactly opposite to \ac{EFTN}). However, when taking the relative difference to $\Lambda$CDM, the modified expansion history and fifth force have opposite effects on $P(k)$ at all scales: Figure \ref{fig:eft_lcdm_enhancement} shows that the modified expansion rate wins at $z=1.0$ and $0.5$ while the fifth force effect dominates at $z=0$.}

\section{Discussions and Conclusions}
\label{sect:discussion}

In this work, we have presented \EFTRAMSES{}, a highly generalised and unified extension of the \ECOSMOG{} N-body simulation optimised for the exploration of the non-linear structure formation under various \ac{MG} and dynamical \ac{DE} scenarios. As cosmology advances into the era of high-precision Stage-IV \ac{LSS} surveys such as \ac{DESI} \citep{DESICollaboration2023}, Euclid \citep{EuclidCollaboration2025}, and LSST \citep{LSSTDESC2012}, relying solely on the background expansion history is no longer adequate to break structural degeneracies. Accessing and modelling the non-linear clustering has become an absolute necessity to thoroughly probe the physical mechanisms responsible for late-time cosmic acceleration.

Our theoretical pipeline originates from the full covariant Horndeski action --- the most general scalar-tensor formulation that preserves second-order field equations to evade Ostrogradsky instabilities. Following the stringent constraints imposed by the GW170817 gravitational wave detection \citep{Abbott2017GW170817}, which required the tensor propagation speed to equal the speed of light ($c_{\text{T}} = c$), higher-order Horndeski derivative couplings ($G_{4X}$ and $G_5$) are effectively eliminated.  This reduction leaves the broader generalised cubic Galileon framework, defined by {three} free functions $G_2(\phi, X)$, $G_3(\phi, X)$, and $G_4(\phi)$ as the most general cosmologically viable cubic-order theory surviving local constraints. The generalised cubic Galileon model is highly compelling because it features a Vainshtein screening mechanism which suppresses the scalar fifth force in high-density environments. However, because the {terms responsible for this screening are driven} by highly non-linear spatial derivative couplings, simulating it presents a severe computational challenge. Rather than hard-coding these complex, model-specific field equations, we embedded the \ac{EFTofDE} directly into our pipeline utilising the parameterisation that this framework offers, especially the phenomenological $\alpha$-basis. 

By mapping the \ac{EFTofDE} framework into a single, unified ``master'' Vainshtein equation, the diverse physical characteristics and non-linear dynamics of the underlying model are completely encapsulated into just three time-dependent, dimensionless background parameters ($\alpha$, $\beta$, and $R_c^2$). This computational strategy bypasses the traditional requirement for developing Poisson solvers; this code can naturally simulate a broad class of scalar-tensor and vector-tensor interactions including {\ac{sDGP}, \ac{nDGP}, \ac{csG}, \ac{cvG}, \ac{GCCG}, and the broader \ac{EFT}} by specifying their background coefficients without altering the underlying non-linear multi-grid solver.

To test this new {code}, we specifically selected the \ac{nDGP}, \ac{csG} and \ac{GCCG} model as our primary benchmarks, because they serve as an ideal, rigorous stress-test. By comparing the non-linear matter power spectrum $P(k)$ generated from 
{\EFTRAMSES{}} against {dependent and independent }numerical baselines {including \ECOSMOG{} and \HICOLA}, we demonstrated {excellent agreement with these latter codes.} 
{In the non-linear regime, for \ac{nDGP} against the old \ECOSMOG{} code and \AREPO{}, deviations $\le 0.5\%$ across all three evaluated redshifts ($z = 1.0, 0.5, 0.0$) up to $k \sim 1.5\,h\,{\rm Mpc}^{-1}$. For \ac{GCCG} in \EFTRAMSES{} against \ECOSMOG{} we observe a sub-percent agreement across all scales at $z = 0$. Finally, \ac{csG} compared to \HICOLA we saw consistent agreement within $1\text{--}2\%$ in the quasi-linear regime ($k \lesssim 1\,h\,{\rm Mpc}^{-1}$), with expected divergence at smaller scales due to \HICOLA's fifth-force approximations.}

{We investigated a persistent large-scale suppression of matter clustering compared to predictions of linear theory, which has been found for all Vainsthein models studied here and grows over time. The same suppression has been found for different codes: \ECOSMOG-based codes such as \ECOSMOG-\textsc{v}, \ECOSMOG-\textsc{eft} and \EFTRAMSES, and independent codes \AREPO{} and \MGGLAM. We found that its origin is the non-linear Vainshtein screening term in Appendix \ref{sec:appendix1}, which explains why the suppression does not appear in simulations where the Vainshtein equation is not explicitly solved, such as \HICOLA. While generally at below $1\%$ level, this shows that linear theory can become inaccurate even on the largest scales of interest for structure formation studies. This may also have an implication for the background expansion rate: looking at the right-hand side of Eq.~\eqref{eq:poisson_eqn_code_units}, the spatial average of $\nabla^2\varphi$ can be nonzero -- in contrast to what we expect for a Poisson equation -- as a result of the non-linear term of Eq.~\eqref{eq:scalar_eom_code_units}. This implies that the non-linear term could contribute to the Friedmann equation, a study we will leave for future works.}

Despite the versatility of this new implementation, several important theoretical limitations remain to be addressed in future work. Most notably, the modified field equations currently implemented within \ac{EFTofDE} rely strictly on the \ac{QSA}. By explicitly dropping the time derivatives ($\delta\dot{\phi}$ and $\delta\ddot{\phi}$) of the scalar field perturbations, the framework essentially treats the scalar field as a massless fluid that adjusts instantaneously to the local matter distribution. While for our pipeline the \ac{QSA} is beneficial for simulating the different models mentioned above, it comes at the theoretical cost {of not being complete \citep[see, e.g.,][for some works considering the importance and potential implications of going beyond the \ac{QSA}]{Barreira:2013,Barreira:2013xea,Winther:2015pta,Moretti:2026axy}}. 

The framework {also} disregards operators responsible for dark energy clustering. Specifically, the current \EFTRAMSES{} equations are entirely blind to the EFT parameter $M_2^4(t)$, which governs the kinetic energy of the scalar perturbations and strictly dictates the sound speed of the dark energy fluid, $c_{\text{s}}$. By dropping this operator, \EFTRAMSES{} implicitly assumes that the scalar field sound speed is comparable to the speed of light ($c_{\text{s}}\sim c$), meaning \ac{DE} cannot cluster on sub-horizon scales. However, for models where $M_2^4$ is present, the sound speed drops significantly ($c_{\text{s}} \ll c$), allowing the dark energy fluid to cluster deeply inside dark matter halos, creating localised $\mathcal{O}(1)$ density perturbations within the dark energy field itself. \EFTRAMSES{} in its current form cannot be used to accurately simulate models where dark energy self-clustering can significantly alter the non-linear growth of structure. Ultimately, addressing these computational restrictions and fully deploying the code outline the next targets for \EFTRAMSES{}. Beyond incorporating full time-dependent field equations to capture \ac{DE} clustering, this pipeline is {well} prepared as the computational counterpart to recent theoretical \ac{DE} reconstruction frameworks \citep[e.g.,][]{Gao:2025}. {See \cite{Ganjoo:2026ugf} for a parallel development with a very similar scope to that of \EFTRAMSES.}

The unified implementation in this work enables robust, high-resolution N-body simulations of \ac{MG} models directly within the \ac{EFTofDE} framework. By focusing purely on relevant \ac{EFT} parameters in these theories, \EFTRAMSES{} significantly enhances our ability to generate precise, non-linear predictions for observational constraints across a diverse range of cosmic scales. 
As part of our future work, we plan to use this code for a detailed exploration of the \ac{EFTofDE} parameter space, coupled with the theoretical reconstruction methods discussed above. It will be deployed to generate full 3D \ac{LSS} and weak lensing maps, precise halo mass functions up to the deeply non-linear regime. Together, these advancements will deliver the numerical instrument {urgently} required to interpret the volume of high-precision data from upcoming Stage-IV surveys and unravel the true physical nature of cosmic acceleration.


\section*{Acknowledgements}

{This work used the DiRAC@Durham facility managed by the Institute for Computational Cosmology on behalf of the STFC DiRAC HPC Facility (https://www.dirac.ac.uk). The equipment was funded by BEIS capital funding via STFC capital grants ST/K00042X/1, ST/P002293/1, ST/R002371/1 and ST/S002502/1, Durham University and STFC operations grant ST/R000832/1. DiRAC is part of the National e-Infrastructure.}

We thank Luis Atayde {and Noemi Frusciante for sharing the matter power spectrum data of the \ac{GCCG} and \ac{QCDM} \ECOSMOG{} simulations} used in Fig.~\ref{GCCGFigure}, and Ashim Sen Gupta and Carola Zanoletti for sharing the power spectrum data of the \ac{csG} \HICOLA{} simulations used in Fig.~\ref{fig:cg_qcdm_master}. We thank Guilherme Brando de Olivera for help to generate the \ac{IC}s used in the \ac{csG} simulations and for discussions.

{NOW, BL and YG are supported by the European Research Council (ERC) Advanced Grant ``UNCA'', under the UKRI's Frontiers Research Guarantee, Grant No.~EP/Z533877/1. SB is supported by the UK Research and Innovation (UKRI) Future Leaders Fellowship (Grant No.~MR/V023381/1 and UKRI2044). YG is additionally supported by a Chinese Scholarship Council visiting PhD studentship. BL is additionally supported by the UK STFC Consolidated Grant No.~ST/X001075/1, and thanks the hosts by the National Astronomical Observatory of China when part of this work was carried out.}

{When this paper was being prepared, another paper with a very similar scope, \cite{Ganjoo:2026ugf}, appeared, introducing the \ECOSMOG-\textsc{EFT} code. Both codes are derived from \ECOSMOG-\textsc{V}. We thank Himanish Ganjoo and Yann Rasera for coordinating the releases of the two codes.}

\section*{Data Availability}

{The power spectrum data and simulation raw data used in this paper are available upon request.}

{\EFTRAMSES{} is publicly available and can be downloaded from \href{https://github.com/nat-woodcock/EFT-Ramses}{this GitHub \EFTRAMSES{} repository}.}




\bibliographystyle{mnras}
\bibliography{example} 



\newpage
\appendix

\section{Investigations of the large-scale mismatch}
\label{sec:appendix1}

As highlighted in Section \ref{subsubsect:nDGP_tests}, the full N-body simulations exhibit a subtle but persistent, {relatively} scale-{in}dependent, discrepancy on large scales ($k \lesssim 0.05~h\mathrm{Mpc}^{-1}$), with respect to the linear theory expectations. {As} this offset was {exactly} replicated by both the legacy \ECOSMOG{} code and the new \EFTRAMSES{} {code}, {and that we have done various checks of the code,} we hypothesised that this does not emerge from the numerical implementation, but rather an inherent physical effect driven by the non-linearities of the Vainshtein mechanism.

\begin{figure}
    \centering
   
    \includegraphics[width=\columnwidth]{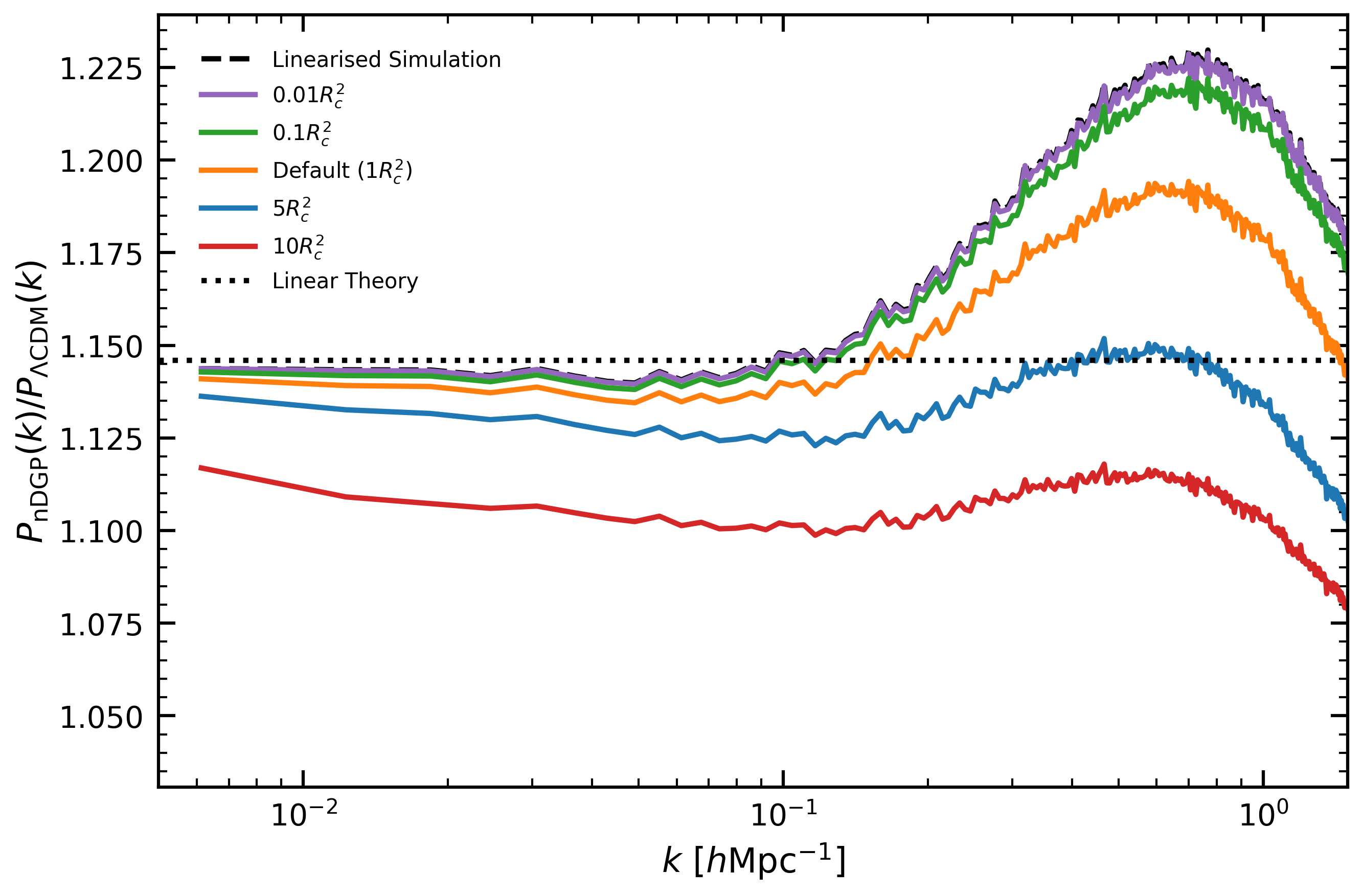}
    \caption{The fractional enhancement of the nDGP matter power spectrum relative to $\Lambda$CDM at $z=0$, extracted from diagnostic simulations without \ac{AMR}. To isolate the physical cause of the large-scale suppression, the coefficient of the non-linear Vainshtein term ($R_c^2$) was artificially scaled. Artificially suppressing this non-linearity by a factor of 0.01 (purple line) perfectly restores the linear theory prediction (dotted black line) on large scales, overlapping identically with a purely linearised simulation (dashed black line). On the other hand, amplifying the non-linearity by factors of 5 (blue line) and 10 (red line) significantly exacerbates the mismatch. This directly confirms that the large-scale deviation from linear theory is an inherent mode-coupling effect driven by the non-linear structure of the Vainshtein equation.}
    \label{fig:rc2_mismatch}
\end{figure}

To demonstrate that this deviation is driven by the non-linear terms in the governing equations, we conducted a series of diagnostic simulations at $z=0$ on a {fixed-resolution (unrefined or no-\ac{AMR})} grid. The Vainshtein screening mechanism is driven by the background function $R_c^2$, By manually multiplying $R_c^2$ by various {numerical} factors, we can systematically modify the strength of these non-linear terms larger or smaller. Figure \ref{fig:rc2_mismatch} presents the resulting matter power spectrum enhancement for scaling factors of $0.01$, $0.1$, $1$ (the default physical value), $5$ and $10$. We also include a ``linearised simulation'' where the scalar field is directly set to the linear theory solution (i.e., the scalar field is not solved but we multiply the Newtonian force with the linear approximation of ${G_{\textrm{eff}}}/{G}$.

We note that manually multiplying $R_c^2$ by a numerical {factor} is not an attempt to physically ``solve'' or correct the large-scale mismatch. Instead, it serves as a diagnostic proof of the origin of the mismatch. We observe that by multiplying $R_c^2$ by a small factor ($0.01$), we reduce the non-linearity of the Vainshtein equation, which predictably restores the linear-theory prediction (dotted grey line) on large scales and is almost identical with the {prediction of the} linearised N-body run (black dashed line). On the other hand, multiplying $R_c^2$ by a large factor ($5$ or $10$) makes the model a lot more non-linear. As shown by the red and blue solid lines in Figure \ref{fig:rc2_mismatch}, this significantly exacerbates the mismatch. This behaviour provides clear, direct evidence that the large-scale deviation from the linear theory is caused by the non-linear terms in the {master} equation. These non-linearities induce mode-coupling that systemically transfers power across scales, leading to a slight suppression of the largest modes. 

\begin{figure}
    \centering
    \includegraphics[width=\columnwidth]{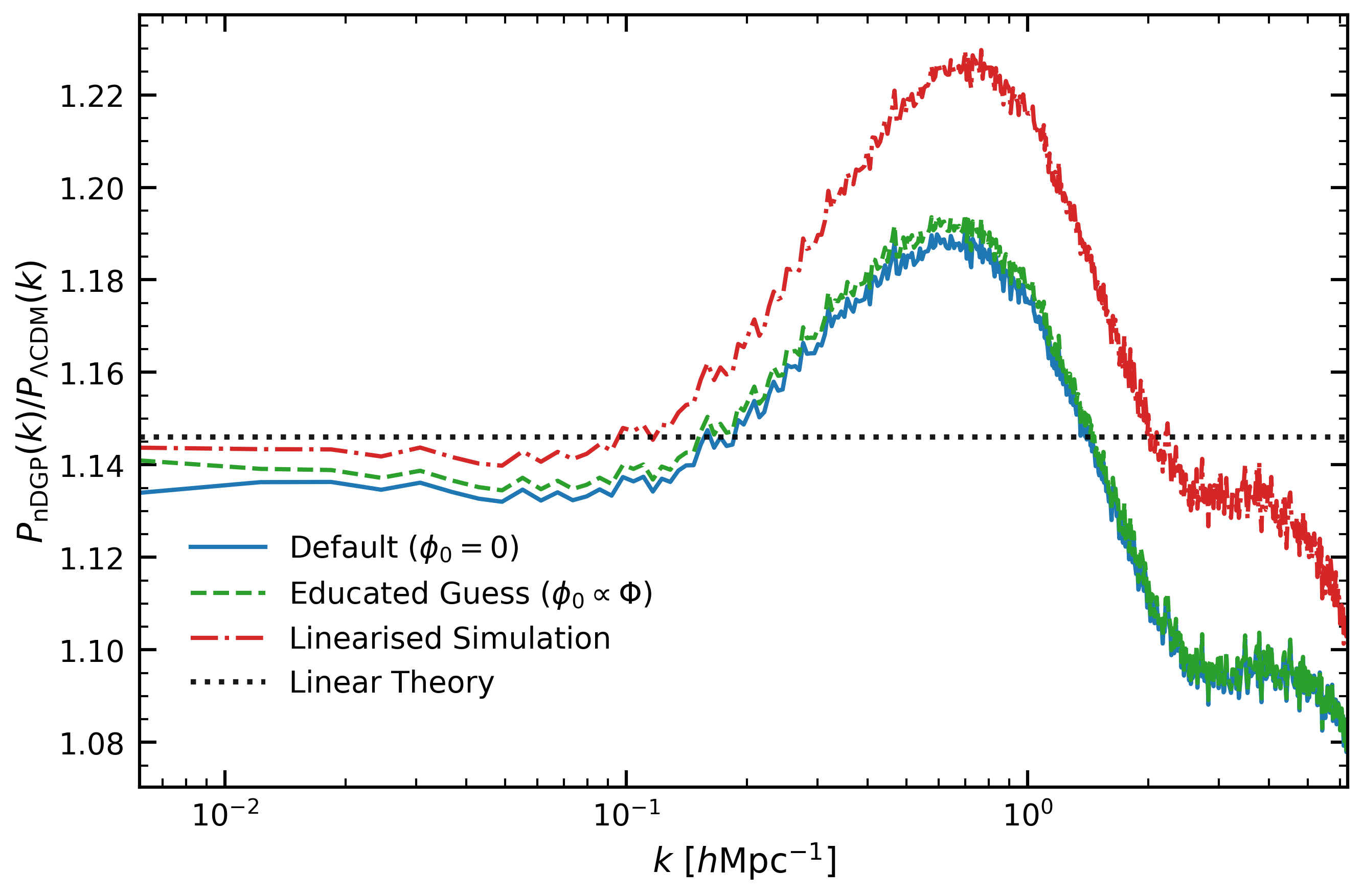} 
    \caption{The fractional enhancement of the matter power spectrum in the nDGP model at $z=0$, extracted from diagnostic simulations run \textit{without} \ac{AMR}. The purple dashed-dotted line represents a purely linearised simulation, which artificially lacks any Vainshtein screening. The blue and green curves represent full non-linear simulations utilising the default initialisation ($\phi_0 = 0$) and an educated initial guess ($\phi_0 \propto \Phi$), respectively. While the base grid resolution is sufficient to trigger the onset of the Vainshtein turnover relative to the linearised run, the lack of \ac{AMR} prevents the solver from reaching the true physical depth seen in our fully refined main results. Furthermore, while the educated guess provides a marginal reduction in the large-scale mismatch on these fixed-resolution simulations, this advantage entirely vanishes once full \ac{AMR} is deployed.}
    \label{fig:no_refinement}
\end{figure}

Furthermore, we investigated the numerical convergence of the simulation 
{against} the impact of the scalar field {initial guess for the multigrid relaxation}. Figure \ref{fig:no_refinement} presents the fractional enhancement extracted from \ac{nDGP} simulations without \ac{AMR}. The blue and green curves represent full non-linear simulations using the default initialisation ($\phi_0 = 0$) and an ``educated guess'' ($\phi_0 = 2\Phi/3\beta$), respectively. {The rationale is that if there is no screening, then the educated guess would be the exact solution to the scalar field, and the large-scale behaviour of this simulation will agree perfectly with the linear-theory prediction. Therefore the educated guess for the scalar field should provide the ``correct'' enhancement that agrees with linear theory, unless the subsequent relaxations damage the long-wavelength modes (to cause the observed mismatch).}

As shown in Figure \ref{fig:no_refinement}, while {using} the educated guess provides a marginal reduction of the mismatch, it is not enough to completely eliminate it. More importantly, we find that this temporary numerical advantage vanishes once full \ac{AMR} is deployed {(cf.~Fig~\ref{fig:ndgp_redshift1})}. Under full high-resolution refinement, the solver robustly converges to the exact same physical solution, preserving both the high-$k$ screening turnover and the inherent large-scale mismatch {largely} independently of its starting state. 

{We carried out several other tests, including changing the number of V-cycles, the depth of the V-cycle (such as completely abandoning multigrid), relaxation sweeps per V-cycle and relaxation convergence criteria. The idea is as follows: if we start with an initial guess that does not have this mismatch, or if we change how the long-wave mode of the error gets reduced during relaxation, and the relaxation iterations always (re)introduce the mismatch, then it is likely inbuilt in the non-linear equation. All of these tests returned the same mismatch: thus the multigrid solver seems always able to couple short-wave and long-wave modes, so that the non-linear terms, which are supposed to be important only on small scales, manage to propagate their effects onto large scales as the relaxation approaches convergence.}

\begin{figure}
    \centering
    \includegraphics[width=\columnwidth]{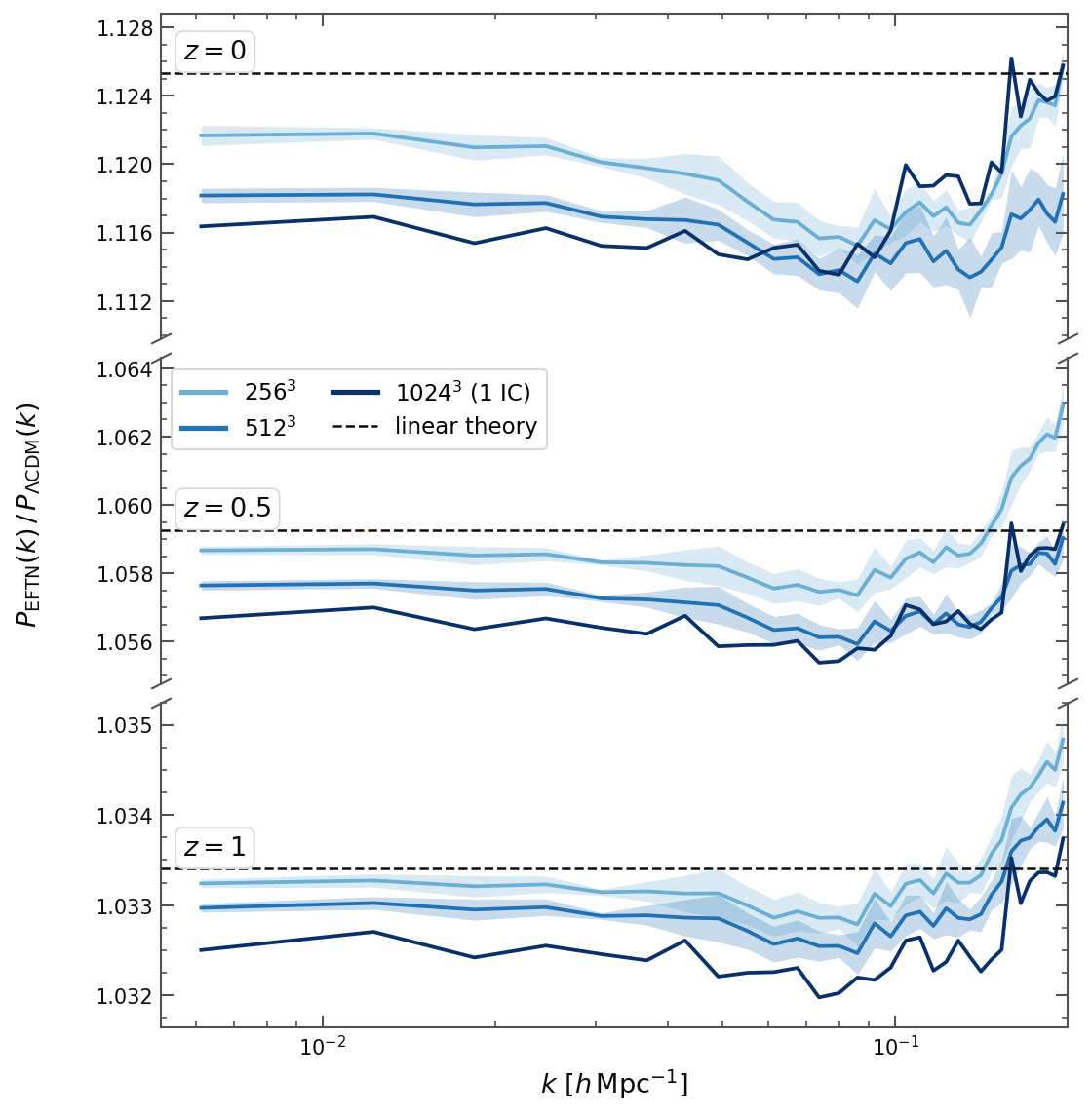} 
    \caption{{Large-scale power spectrum enhancement $P_{\rm EFTN}(k)/P_{\Lambda\rm CDM}(k)$ for the \ac{EFTN} model across three redshifts ($z=0, 0.5, 1.0$), demonstrating the recovery of the spectral enhancement towards the linear theory line. Light blue and mid-blue solid lines show the ensemble mean over 5 independent initial realisations for $256^3$ and $512^3$ particles with $L_{\rm box}=1024h^{-1}$Mpc, respectively, with shaded region indicating the 95\% confidence intervals across the 5 realisations. The dark blue solid line shows the single-realisation $1024^3$ simulation from the main text as a high-resolution reference, and the horizontal black dashed lines represent the prediction of linear perturbation theory. Coarsening the grid resolution smooths out local density peaks and suppresses spatial derivative cross-terms, rendering the discrete equation less non-linear and systematically approaching back to the linear-theory limit on large scales.}}
    \label{fig:eftnav}
\end{figure}

{To further validate the hypothesis, in Figure \ref{fig:eftnav} we have compared the power spectrum enhancement $P_{\rm EFTN}(k)/P_{\Lambda\rm CDM}(k)$ across three particle resolutions $256^3$, $512^3$, and $1024^3$ particles at a fixed box size of $L_{\rm box} = 1024\,h^{-1}\,{\rm Mpc}$. Given that individual low-$k$ Fourier modes are subject to cosmic variance in finite volumes, for the two relatively cheap lower-resolution simulations we have simulated an ensemble of five independent realisations each. The shaded regions indicate the 95\% confidence intervals across this five-realisation ensemble. 
}
{At all three redshifts shown, a clear resolution trend emerges, with coarser simulations aligning more closely to linear theory at the low $k$ where the large-scale mismatch was seen. This trend reflects the numerical behaviour of the Vainshtein term, $\mathcal{O}_{\rm NL}(\varphi) \equiv (\nabla^2\varphi)^2 - \nabla_i\nabla_j\varphi\nabla^i\nabla^j\varphi$: coarsening the grid smooths localised density peaks over $8$-$64$ times larger cell volumes and damps discrete finite difference curvatures in the non-linear terms ($\propto k^4$) relative to the Laplacian ($\propto k^2$) in the master Vainshtein equation. In effect, lowering the particle resolution removes/suppresses the short-wavelength (high-$k$) modes and steep density gradients, which inherently reduces the non-linearity of the equations.}

\begin{figure}
    \centering
    \includegraphics[width=\columnwidth]{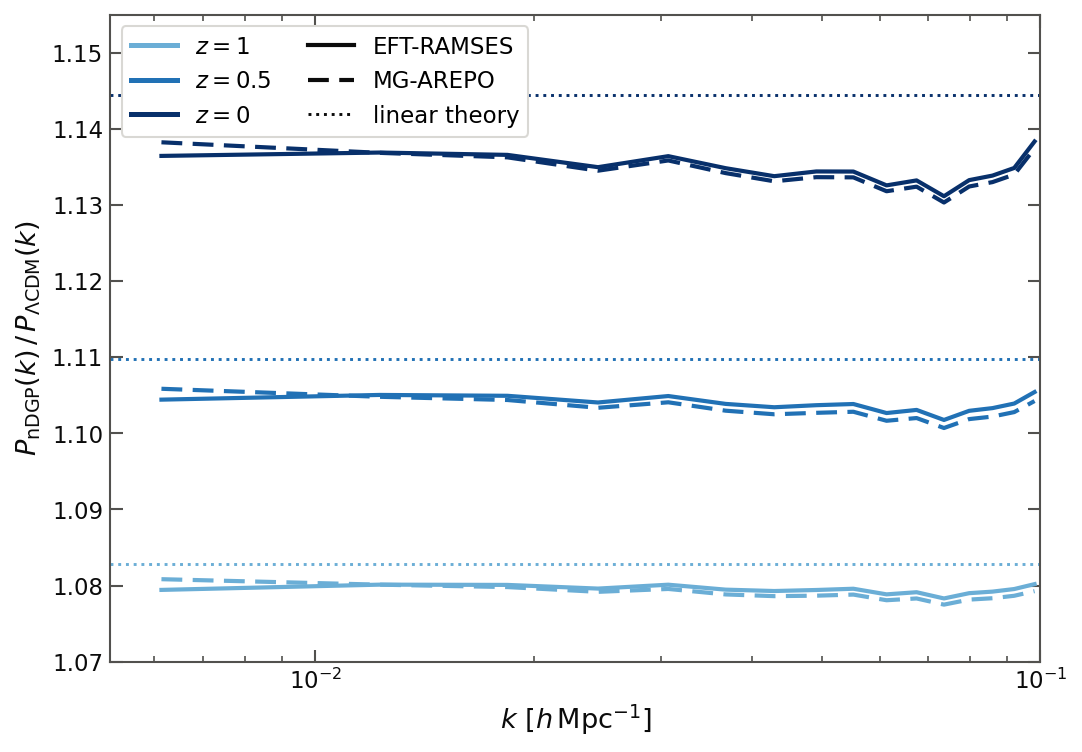}
    \caption{{Matter power spectrum enhancement $P_{\rm nDGP}(k)/P_{\Lambda\rm CDM}(k)$ on large scales ($0.005 \le k \le 0.1\,h\,{\rm Mpc}^{-1}$) for the nDGP model across three redshifts: $z=1.0$ (light blue), $z=0.5$ (mid blue), and $z=0.0$ (dark navy) with $N_{\rm p} = 1024^3$ particles. Solid lines denote \EFTRAMSES{}, dashed lines denote the independent moving-mesh code \AREPO{}, and horizontal dotted lines mark the linear perturbation theory predictions. Both codes agree to within $\lesssim 0.2\%$ across all resolved large-scale modes. Crucially, \AREPO{} reproduces the exact same systematic suppression relative to linear perturbation theory at late times, demonstrating that this mismatch is not driven by the difference in the numerical implementation.}}
    \label{fig:ndgplarge}
\end{figure}

{Finally, as a clean test of whether the large-scale mismatch could have been caused by the specific numerical implementations of \EFTRAMSES, we compared its predictions against the moving-mesh code \AREPO{} \citep{Hernandez-Aguayo2021}, for the \ac{nDGP} model. Figure \ref{fig:ndgplarge} shows the large-scale matter power spectrum enhancement $P_{\rm nDGP}(k)/P_{\Lambda\rm CDM}(k)$ on linear scales ($0.005 \le k \le 0.1\,h\,{\rm Mpc}^{-1}$) from \EFTRAMSES{} and \AREPO{} at three redshifts ($z = 1.0, 0.5, 0.0$). The two independent codes agree very well at these scales and all 3 redshifts, with \AREPO{} producing the same sub-percent mismatch with linear theory prediction. This serves as further confirmation that the large-scale mismatch is an intrinsic consequence of the non-linear Vainshtein screening term in the scalar \ac{EoM}.}


\bsp	
\label{lastpage}
\end{document}